\documentclass[pra,aps,reprint,superscriptaddress,amsmath,amssymb]{revtex4-2}
\usepackage{graphicx,bm,mathptmx,physics,color,multirow,array}
\usepackage[colorlinks=true, allcolors=blue]{hyperref}
\usepackage[normalem]{ulem}
\def\nih#1{{\color{black}#1}}

\begin{document}

\title{Subcycle resolved strong-field tunneling ionization: \\Identification of magnetic dipole and electric quadrupole effects}

\author{Xiaodan~Mao}
\affiliation{State Key Laboratory of Precision Spectroscopy, East China Normal University, Shanghai 200241, China}
\author{Hongcheng~Ni}
\email{hcni@lps.ecnu.edu.cn}
\affiliation{State Key Laboratory of Precision Spectroscopy, East China Normal University, Shanghai 200241, China}
\affiliation{Institute for Theoretical Physics, Vienna University of Technology, 1040 Vienna, Austria}
\affiliation{Collaborative Innovation Center of Extreme Optics, Shanxi University, Taiyuan, Shanxi 030006, China}
\affiliation{NYU-ECNU Joint Institute of Physics, New York University at Shanghai, Shanghai 200062, China}
\author{Xiaochun~Gong}
\affiliation{State Key Laboratory of Precision Spectroscopy, East China Normal University, Shanghai 200241, China}
\affiliation{Collaborative Innovation Center of Extreme Optics, Shanxi University, Taiyuan, Shanxi 030006, China}
\author{Joachim~Burgdörfer}
\affiliation{Institute for Theoretical Physics, Vienna University of Technology, 1040 Vienna, Austria}
\author{Jian~Wu}
\email{jwu@phy.ecnu.edu.cn}
\affiliation{State Key Laboratory of Precision Spectroscopy, East China Normal University, Shanghai 200241, China}
\affiliation{Collaborative Innovation Center of Extreme Optics, Shanxi University, Taiyuan, Shanxi 030006, China}
\affiliation{NYU-ECNU Joint Institute of Physics, New York University at Shanghai, Shanghai 200062, China}
\affiliation{CAS Center for Excellence in Ultra-intense Laser Science, Shanghai 201800, China}

\begin{abstract}
Interaction of a strong laser pulse with matter transfers not only energy but also linear momentum of the photons. Recent experimental advances have made it possible to detect the small amount of linear momentum delivered to the photoelectrons in strong-field ionization of atoms. Linear momentum transfer is a unique signature of the laser-atom interaction beyond its dipolar limit. Here, we present a decomposition of the subcycle time-resolved linear momentum transfer in term of its multipolar components. We show that the magnetic dipole contribution dominates the linear momentum transfer during the dynamical tunneling process while the post-ionization longitudinal momentum transfer in the field-driven motion of the electron in the continuum is primarily governed by the electric quadrupole interaction. Alternatively, exploiting the radiation gauge, we identify nondipole momentum transfer effects that scale either linearly or quadratically with the coupling to the laser field. The present results provide detailed insights into the physical mechanisms underlying the subcycle linear momentum transfer induced by nondipole effects.
\end{abstract}

\maketitle

\section{Introduction}

Tunneling ionization is the first step of many strong-field phenomena and is one of the cornerstones of strong-field and attosecond physics. Tunneling ionization from atoms and molecules is typically described within the electric dipole approximation, where the laser field is considered homogeneous along its propagation direction, thereby neglecting the photon momentum and the field retardation \cite{Reiss2014}. The dipole approximation holds well for typical parameters of experimental table-top laser setups. In turn, nondipole effects, which result in asymmetric photoelectron momentum distributions about zero along the field propagation direction, are typically very small.

With advances in the detection technology, such weak nondipole effects have recently become accessible \cite{Peng2020,Keller2021}. In 2011, \emph{Smeenk et al.}~\cite{Smeenk2011} experimentally observed the linear momentum transfer for tunneling ionization of argon and neon. Since then, nondipole tunneling effects have attracted considerable attention. It has been found to cause a negative shift in the photoelectron momentum distribution for linear polarization due to its interplay with the Coulomb field \cite{Ludwig2014}. Remarkably, for nonsequential double ionization, the sum of the nondipole momentum shifts of the two electrons is opposite to the case of single ionization and considerably larger \cite{Sun2020}. The nondipole momentum shift follows the prediction of the classical model for low photoelectron energies \cite{Hartung2019}, while for high-order above-threshold ionization, the nondipole momentum shift is found to be substantially modified by large-angle rescattering of the electron \cite{Brennecke2018,Lin2022}. Nondipole effects also play an observable role during the under-barrier tunneling ionization \cite{He2022,Klaiber2022}. Moreover, nondipole effects induce a modification to the ponderomotive potential thereby shifting the center of the energy rings in the above-threshold ionization \cite{Brennecke2021,Lund2021,Lin2022sa}. In 2019, the pioneering experimental work of \emph{Willenberg et al.}~\cite{Willenberg2019} studied the subcycle linear momentum transfer with an attoclock protocol \cite{Eckle2008a,Eckle2008b}, and found a counterintuitive local minimum of the transferred momentum at the peak of the laser electric field. Based on earlier theories of nondipole effects \cite{Sommerfeld1930,Klaiber2013,Liu2013,Chelkowski2014,Chelkowski2017,Wang2017,He2017,Jensen2020,Madsen2022}, we have subsequently developed the theory of subcycle linear momentum transfer \cite{Ni2020}, which accounts for the interplay between nondipole and nonadiabatic tunneling effects on the sub-optical-cycle time scale.

Different contributions to the linear momentum transfer by nondipole effects can be distinguished. In the time domain, the momentum transfer $\langle p_z \rangle$ along the laser propagation direction ($\hat{z}$) can be partitioned as \cite{Eicke2020,Ni2020}
\begin{equation}
\langle p_z \rangle = \langle v_z \rangle + \Delta E/c
\end{equation}
with $\langle v_z \rangle$ the momentum transfer due to tunneling ionization and $\Delta E /c$ the contribution due to the subsequent continuum motion after tunneling with $\Delta E$ denoting the energy gain during the continuum excursion of the electron and $c$ representing the speed of light. An alternative partitioning can be achieved in the energy domain by exploring the influence of different terms of nondipole Hamiltonian to a given order $\alpha$ ($\alpha$ is the fine structure constant) \cite{Anzaki2018,Bachau2019,Liang2018} on the linear momentum transfer. Recently, the time-integrated contributions of the electric field gradient and the magnetic field to the final momentum distribution in tunneling ionization have been studied \cite{Hartung2021}.

In the present article, we aim at combining the time domain and spectral domain analysis in order to pinpoint the influence of individual nondipole terms in the Hamiltonian on the subcycle time-resolved momentum transfer within a typical attoclock protocol. Using an elliptically polarized laser pulse, the electrons emitted at different instances of time within an optical cycle are emitted into different directions and backscattering is suppressed, which facilitates the subcycle time resolution of nondipole-induced momentum transfer. In the present study, we consider two frequently employed gauges, the multipole gauge and the radiation gauge. The multipole gauge allows to identify the contributions of the magnetic dipole and electric quadrupole interactions to leading order $1/c$ in the retardation. The radiation gauge, on the other hand, reveals separate contributions linear and quadratic in the laser field to the same order $1/c$. While in exact calculations the results should be invariant under changes of the gauge, approximate treatments may be sensitive to the choice of gauge. A well-known case in point are dipole transitions in weak-field photoionization. Depending on the gauge employed (length, velocity, accelerations), different spatial regions of the wave functions can be sensitively probed. Here, we extend such an analysis to the nondipole strong-field regime. We show that the electric quadrupole (E$_2$) coupling primarily probes the field-driven continuum motion which, overall, provides the dominant contribution to the linear momentum transfer while the tunneling process is controlled by the magnetic dipole (M$_1$) coupling, which provides a minor correction for the present parameter settings.

The article is organized as follows. In Sec.~II, we briefly review the theoretical framework within which we treat the partitioning of the leading-order nondipole corrections in the Hamiltonian for atom-laser interaction. In Sec.~III, the subcycle contributions originating from each nondipole Hamiltonian term are analyzed and approximate analytic expressions within the framework of the strong-field approximation (SFA) are given. We present results for the subcycle contributions of the different nondipole interactions in Sec.~IV. The time integral momentum transfer as a function of the laser parameters is explored in Sec.~V. Conclusions are given in Sec.~VI. Atomic units are used throughout unless noted otherwise.

\section{Theoretical framework}

\subsection{The nondipole Hamiltonian}

The light wave of a laser in vacuum is a transverse wave with polarization perpendicular to the propagation direction ($\hat{\bm{k}}$) along which the linear photon momentum $k=\omega/c$ ($\omega$: laser angular frequency) points. In the following,  $z$ is chosen along the laser propagation direction and $(x,y)$ represents the laser polarization plane. For inclusion of nondipole effects into light-matter interactions, the spatio-temporal dependence of the vector potential of the laser field $\bm{A}(\bm{r},t) = \bm{A}(t-z/c)$ needs to be considered. Focusing in the following on the Hamiltonian operator of an (effective) one-electron atom with atomic potential $V(\bm{r})$, the minimal-coupling Hamiltonian in the radiation gauge is given by
\begin{equation}
H = \frac{1}{2}\left[\bm{p}+\bm{A}(\bm{r},t)\right]^2 + V(\bm{r}).
\label{eq:H}
\end{equation}
Since nondipole retardation effects $\sim1/c$ are the focus of our present study, we have absorbed other $1/c$ factors, in particular those originating from the coupling between charged particles and the radiation field in terms of the fine structure constant $\alpha$ ($\alpha=1/c$ in atomic units), into the amplitude of the effective vector potential $\bm{A}(\bm{r},t)$. The minimal-coupling Hamiltonian [Eq.~\eqref{eq:H}] can be decomposed into terms of first and second order couplings to the field
\begin{equation}
H = \frac{1}{2}p^2 + \underbrace{\bm{p}\cdot\bm{A}(t-z/c)}_\text{linear in field} + \underbrace{\frac{1}{2}A^2(t-z/c)}_\text{quadratic in field} + V(\bm{r}).
\end{equation}

Expanding now the vector potential to first order in $c^{-1}$, representing field retardation or nondipole effects to lowest order,
\begin{equation}
\bm{A}(t-z/c)\approx\bm{A}(t)-\frac{z}{c}\dot{\bm{A}}(t)=\bm{A}(t)+\frac{z}{c}\bm{F}(t),
\end{equation}
with the electric field $\bm{F}(t)=-\dot{\bm{A}}(t)$, yields the effective Hamiltonian $H_{\rm ND}$ including nondipole effects,
\begin{equation}
H_{\rm ND}=\frac{1}{2}\left[\bm{p}+\bm{A}(t)\right]^2+V(\bm{r})+\underbrace{\frac{z}{c}\bm{F}(t)\cdot\bm{p}}_{\mathrm{F}_1\text{ term}}+\underbrace{\frac{z}{c}\bm{F}(t)\cdot\bm{A}(t)}_{\mathrm{F}_2\text{ term}},
\label{eq:H_F1F2}
\end{equation}
where nondipole retardation effects ($\sim1/c$) appear both linear $\left(z/c\right)[\bm{F}(t)\cdot\bm{p}]$ (the F$_1$ term) and quadratic $\left(z/c\right)[\bm{F}(t)\cdot\bm{A}(t)]$ (the F$_2$ term) in the laser field.

Applying now the Powers-Zienau-Wolley gauge transformation
\begin{equation}
\Psi'(\bm{r},t)=e^{i\Lambda_{\rm ND}(\bm{r},t)}\Psi(\bm{r},t)
\label{eq:GT_1}
\end{equation}
with the nondipole gauge phase
\begin{equation}
\Lambda_{\rm ND}(\bm{r},t)=\int_{0}^{1} \bm{r}\cdot \bm{A}(t-\lambda \frac{z}{c})d\lambda,
\label{eq:GT_2}
\end{equation}
the transformed Hamiltonian
\begin{align}
H_{\rm ND}' &= e^{i\Lambda_{\rm ND}}\left(H_{\rm ND}-i\frac{\partial}{\partial t}\right)e^{-i\Lambda_{\rm ND}} \nonumber \\
&= e^{i\Lambda_{\rm ND}}H_{\rm ND}e^{-i\Lambda_{\rm ND}} -\frac{\partial}{\partial t}\Lambda_{\rm ND}
\label{eq:GT_3}
\end{align}
becomes \cite{Anzaki2018,Liang2018}
\begin{equation}
H_{\rm ND}' = \frac{\bm{p}^2}{2} + V(\bm{r}) + \underbrace{\bm{r}\cdot\bm{F}(t)}_{\mathrm{E}_1\text{ term}} +
\underbrace{\frac{1}{2c}\bm{L}\cdot\bm{B}(t)}_{\mathrm{M}_1\text{ term}}
\underbrace{-\frac{z}{2c}\left[\bm{r}\cdot\dot{\bm{F}}(t)\right]}_{\mathrm{E}_2\text{ term}}
\label{eq:H_E2M1}
\end{equation}
with angular momentum $\bm{L} = \bm{r}\times\bm{p}$ and the effective magnetic field $\bm{B}(t)=\bm{\hat k}\times\bm{F}(t)$. Eq.~\eqref{eq:H_E2M1} represents the multipole expansion of the Hamiltonian, $H_{\rm ND}'$, including the electric (${\rm E}_1$) and magnetic dipole (${\rm M}_1$) and electric quadrupole (${\rm E}_2$) contributions. The properties of these particle--radiation-field interaction terms are well understood within the framework of lowest order perturbation theory. We explore in the following the role of the terms beyond the standard dipole (${\rm E}_1$) approximation in strong-field ionization and, in particular, in nonadiabatic tunneling ionization.

To further pinpoint the influence of individual nondipole terms separately, we will consider also reduced Hamiltonians where only selected nondipole terms are kept,
\begin{align}
H_{{\rm M}_1} &= \frac{\bm{p}^2}{2} + V(\bm{r}) + \bm{r}\cdot\bm{F}(t) + \frac{1}{2c}\bm{L}\cdot\bm{B}(t),
\label{eq:M1} \\
H_{{\rm E}_2} &= \frac{\bm{p}^2}{2} + V(\bm{r}) + \bm{r}\cdot\bm{F}(t) - \frac{z}{2c}\left[\bm{r}\cdot\dot{\bm{F}}(t)\right].
\label{eq:E2}
\end{align}
An analogous decomposition can also be performed for the Hamiltonian in the radiation gauge [Eq.~\eqref{eq:H_F1F2}]. Accordingly, we will consider reduced Hamiltonians which include nondipole terms ${\rm F}_1$ or ${\rm F}_2$,
\begin{align}
H_{{\rm F}_1} &= \frac{1}{2}\left[\bm{p}+\bm{A}(t)\right]^2+V(\bm{r})+\frac{z}{c}\bm{F}(t)\cdot\bm{p},
\label{eq:F1} \\
H_{{\rm F}_2} &= \frac{1}{2}\left[\bm{p}+\bm{A}(t)\right]^2+V(\bm{r})+\frac{z}{c}\bm{F}(t)\cdot\bm{A}(t).
\label{eq:F2}
\end{align}
For each of these reduced Hamiltonians [Eqs.~(\ref{eq:M1}--\ref{eq:F2})], a suitable gauge transformation analogous to Eqs.~(\ref{eq:GT_1}--\ref{eq:GT_3}) will be performed.

\subsection{Nondipole strong-field approximation}

We focus in the following on the momentum transfer $\langle p_z\rangle$ along the laser propagation direction as the characteristic signature of nondipole effects in strong-field ionization. In a recent publication \cite{Ni2020}, we have shown by a comparison between the full numerical solution using the time-dependent Schr\"odinger equation (TDSE) and the SFA that for strong-field ionization within an attoclock scenario with elliptically or circularly polarized radiation, the atomic (or Coulomb) potential $V(\rm{r})$ [Eqs.~(\ref{eq:H_F1F2},\ref{eq:H_E2M1})] has only negligible influence on $\langle p_z\rangle$ as rescattering and laser-Coulomb coupling are suppressed. We have verified that the atomic force field has only a negligible influence also on the decomposition into individual nondipole contributions (see Appendix~A). Consequently, we investigate the contributions of the different nondipole terms to $\langle p_z\rangle$ within the SFA. Moreover, by applying a saddle-point approximation (SPA), we can extract approximate analytic expressions for these nondipole contributions, thereby providing a transparent physical picture of the subcycle contributions of different nondipole interactions.

Within the SPA including nondipole corrections, labeled ndSPA in the following, the triply differential transition rate to the final state with momentum $\bm{p}$ is given by \cite{Gribakin1997,Kjeldsen2006,Milo2006}
\begin{equation}
W_\mathrm{ndSPA}(\bm{p})=|\ddot{S}|^{-\alpha_Z}\exp\{2\mathrm{Im}S\},
\label{eq:rate}
\end{equation}
where $\alpha_Z = 1+Z/\sqrt{2I_p}$ with $Z$ the asymptotic charge of the remaining ion and $I_p$ the ionization potential, $S$ denotes the nondipole action,
\begin{equation}
S=\int_{t_s}^{t_r}\left( H_{\rm ND} +I_p\right) dt,
\end{equation}
where $H_{\rm ND}$ presents the Hamiltonian including nondipole terms in various gauges but not the Coulomb potential, and
\begin{equation}
\dot{S}=H_{\rm ND}+I_p=0
\label{eq:SPE}
\end{equation}
is the corresponding saddle-point equation. Eq.~\eqref{eq:SPE} determines the complex saddle-point time $t_s=t_r+it_i$. The real part of $t_s$ represents the ionization time $t_r$, the instant the tunneling electron becomes free, while the imaginary part $t_i$ is related to the tunneling rate. For the numerical results presented in the following, we mostly use $\alpha_Z=1$. \nih{As shown in Appendix~F, we} have verified that its exact value only plays a minor role and does not change the conclusions because the tunneling dynamics is dominated by the exponential factor [Eq.~\eqref{eq:rate}].

For the determination of the linear momentum transfer at the tunneling exit $\langle v_z \rangle$ and the asymptotic momentum transfer $\langle p_z \rangle$, it is advantageous to make a coordinate transformation $(p_x, p_y, p_z)\rightarrow(t_r, k_\perp, p_z)$. With the proper choice of the auxiliary momentum $k_\perp$ (the subscript $\perp$ denotes variables in the laser polarization plane), one of the two saddle-point equations, $\Im\dot{S}=0$, can be automatically fulfilled. Consequently, the search for the saddle-point time $t_s$ in the complex plane can be reduced to the search along one axis, thereby greatly reducing the computational cost. The choice of $k_\perp$ will depend on $H_{\rm ND}$ as specified below. The auxiliary transverse momentum $k_\perp$ turns out to be closely related to the transverse momentum $v_\perp$ at the tunnel exit. The transition rate $\tilde{w}$ as a function of $t_r$, $k_\perp$, and $p_z$ includes the Jacobian of the coordinate transformation
\begin{equation}
\tilde{w}_\mathrm{ndSPA}(t_r,k_\perp,p_z)=\left|\mathrm{det}\frac{\partial(p_x,p_y,p_z)}{\partial(t_r,k_\perp,p_z)}\right|W_\mathrm{ndSPA}(\bm{p}).
\label{eq:WndSPAjac}
\end{equation}
Eq.~\eqref{eq:WndSPAjac} can be extended to the case where multipole saddle points contribute to the emission to the same final momentum $\bm{p}$. In this case, the transition rate $W_{\rm ndSPA}$ is given by
\begin{equation}
W_\mathrm{ndSPA}(\bm{p})=\left|\sum_nM_{\bm{p}}^{(n)}\right|^2,
\end{equation}
where $M_{\bm{p}}^{(n)}=|\ddot{S}^{(n)}|^{-\alpha_Z/2}\exp\{-iS^{(n)}\}$. For the numerical examples presented below we find one saddle point to be dominant and the influence of such coherent superpositions to be negligible.

The expectation value of the subcycle time-resolved linear momentum transfer at the tunnel exit follows as
\begin{equation}
\langle v_z(t_r) \rangle=\frac{\int \mathrm{d}k_\perp \mathrm{d}p_z\, v_z(t_r,k_\perp,p_z)\, \tilde{w}_\mathrm{ndSPA}(t_r,k_\perp,p_z)}{\int \mathrm{d}k_\perp \mathrm{d}p_z\, \tilde{w}_\mathrm{ndSPA}(t_r,k_\perp,p_z)},
\label{eq:partial}
\end{equation}
and consequently the asymptotic longitudinal momentum $\langle p_z(t_r) \rangle$ as \cite{Eicke2020,Ni2020}
\begin{equation}
\langle p_z (t_r) \rangle =\langle v_z(t_r) \rangle + \frac{\Delta E}{c},
\label{eq:pz_all}
\end{equation}
with $\Delta E = \left(\langle p_\perp^2\rangle-\langle v_\perp^2\rangle \right) /2$. More generally, the coordinate transformation $(p_x, p_y, p_z)\rightarrow(t_r, k_\perp, p_z)$ allows to determine also other variables of interest as a function of the tunneling ionization time $t_r$, thereby unraveling information on quantum dynamics along the time axis. Explicit expressions for the auxiliary transverse momentum $k_\perp$ and $v_z (t_r, k_\perp, p_z)$ appearing in the integrand [Eq.~\eqref{eq:partial}] depend on the specific form of the nondipole Hamiltonian under consideration and will be given in Sec.~III [Eqs.~(\ref{eq:k_full},\ref{eq:vz_lab})] as well as in the Appendices~B to E. Results obtained using Eq.~\eqref{eq:WndSPAjac} are labeled ndSPA in what follows.

\subsection{Nonadiabatic expansion}

Elementary strong-field tunneling theory often referred to as the simple man's model relies on the adiabatic limit of vanishingly small Keldysh parameter $\gamma$ in which the strong field is treated as quasistatic. Consequently, the tunneling barrier is considered to be time independent and tunneling transitions are ``horizontal'' and energy conserving. For small but finite Keldysh parameter $\gamma\approx\omega t_i$ nonadiabatic corrections to the quasistatic tunneling process should be accounted for. To determine lowest-order nonadiabatic corrections, we expand the vector potential $\bm{A}(t_s)=\bm{A}(t_r+it_i)$ up to the second order in $t_i$ \cite{Goreslavski1996,Shvetsov2003,Goreslavski2004,Frolov2017,Ni2018,Ni2020,Ma21},
\begin{equation}
\bm{A}(t_r+it_i) \approx \bm{A}(t_r) - it_i\bm{F}(t_r)+\frac{1}{2}t^2_i\dot{\bm{F}}(t_r).
\label{eq:A}
\end{equation}

Restricting the inclusion of nonadiabatic corrections to this order [Eq.~\eqref{eq:A}] allows the fully analytic evaluation of ionization rate [Eq.~\eqref{eq:WndSPAjac}]. The results obtained in this approach are hereafter labeled as nondipole saddle-point approximation with nonadiabatic expansion (ndSPANE). We note that this inclusion of nonadiabatic effects is sometimes referred to as the adiabatic expansion \cite{Ni2018,Ma21}. It should be further noted that the $k_\perp$ determined through Eq.~\eqref{eq:SPE} may differ from each other when the ndSPA and ndSPANE are used. These deviations do not significantly affect the numerical results.

\section{Evaluation of the nondipole strong-field approximation}

In this section, we apply the ndSPA and the ndSPANE, as outlined above to the subcycle linear momentum transfer with the goal to identify the contributions from individual nondipole terms in the Hamiltonian [Eqs.~\eqref{eq:H_F1F2} and \eqref{eq:H_E2M1}]. We first demonstrate the evaluation for full $H_{\rm ND}$ in the radiation gauge [Eq.~\eqref{eq:H_F1F2}]. The analogous evaluation for the alternative decompositions of $H_{\rm ND}$ are given in the Appendix.

We first apply a gauge transformation to Eq.~\eqref{eq:H_F1F2} with gauge function
\begin{equation}
\Lambda = -\frac{z}{c}\left[\left(\bm{p}\cdot\bm{A}(t)+\frac{1}{2}A^2(t)\right)\right],
\label{eq:GT_all}
\end{equation}
which results in \cite{Brennecke2018,Hartung2019,Ni2020}
\begin{align}
H_{\rm ND}' =& \frac{1}{2}\left[\bm{p}+\bm{A}(t)+\frac{\hat{\bm{e}}_z}{c}\left(\bm{p}\cdot\bm{A}(t)+\frac{1}{2}A^2(t) \right)\right]^2 \nonumber\\
&+ V\left(\bm{r}-\frac{z}{c}\bm{A}(t)\right).
\label{eq:H_Total_new}
\end{align}
This gauge transformation implies a time-dependent shift of the origin in the nondipole frame similar to the Kramers-Henneberger gauge. Even when $V(\bm{r})$ is eventually neglected in the SFA, such a frame transformation is needed to retrieve quantities in the lab frame from results obtained in this nondipole frame. The corresponding action entering the ndSPA is
\begin{equation}
S=\int_{t_s}^{t_r}\left\{\frac{1}{2}\left[\bm{p}+\bm{A}(t)+\frac{\hat{\bm{e}}_z}{c}\left(\bm{p}\cdot\bm{A}(t)+\frac{1}{2}A^2(t) \right)\right]^2 + I_p \right\}dt,
\end{equation}
and the saddle-point equation reads
\begin{equation}
\frac{1}{2}\left[\bm{p}+\bm{A}(t_s)+\frac{\hat{\bm{e}}_z}{c}\left(\bm{p}\cdot\bm{A}(t_s)+\frac{1}{2}A^2(t_s) \right)\right]^2+ I_p = 0.
\label{eq:S_Total}
\end{equation}

With the substitution $\bm{k}_\perp = \bm{p}_\nih{\perp}+{\rm Re}\bm{A}(t_s)$, the imaginary part of the saddle-point equation becomes
\begin{equation}
\left(1+\frac{p_z}{c}\right)\left[k_x\Im A_x(t_s)+k_y\Im A_y(t_s)\right]=0.
\end{equation}
Therefore, by choosing the auxiliary perpendicular momentum $k_\perp$ in the polarization plane as
\begin{equation}
k_\perp = \left(\bm{p}_\nih{\perp}+\Re \bm{A}(t_s) \right)\cdot\frac{\Im A_y(t_s)\hat{\bm{e}}_x - \Im A_x(t_s)\hat{\bm{e}}_y}{\sqrt{\left(\Im A_x(t_s) \right)^2+\left(\Im A_y(t_s) \right)^2}},
\label{eq:k_full}
\end{equation}
the imaginary part of the saddle-point equation [Eq.~\eqref{eq:S_Total}] is automatically fulfilled for arbitrary pulse shapes. The relation between the auxiliary momentum $k_\perp$ [Eq.~\eqref{eq:k_full}] and the transverse momentum at the tunnel exit $v_\perp$ will be explored below. Expressing the momentum differential ionization probability $W_{\rm ndSPA}(\bm{p})$ [Eq.~\eqref{eq:rate}] in terms of the coordinates ($t_r$, $k_\perp$, $p_z$), the time-resolved initial linear momentum $\langle v_z(t_r) \rangle$ is obtained by evaluating Eqs.~\eqref{eq:WndSPAjac} and \eqref{eq:partial}, and the final linear momentum $\langle p_z(t_r) \rangle$ follows from Eq.~\eqref{eq:pz_all}.

Analytic expressions for $\langle v_z \rangle$ and $\langle p_z \rangle$ can be obtained when keeping only lowest-order nonadiabatic corrections using ndSPANE. Inserting $\bm{A}(t_r+it_i)$ [Eq.~\eqref{eq:A}] into the saddle-point equation [Eq.~\eqref{eq:SPE}] and keeping terms up to second order in $t_i$ results in the relationship
\begin{equation}
\bm{k}_\perp\cdot\bm{F}(t_r)=\left[\bm{p}_\perp +\bm{A}(t_r) \right]\cdot\bm{F}(t_r)=0,
\label{eq:kF}
\end{equation}
i.e., the transverse momentum $\bm{k}_\perp$ is orthogonal to the attoclock field at the instant of tunneling ionization $t_r$. Eq.~\eqref{eq:kF} defines the temporal location $t_r$ of the tunnel exit separating tunneling from continuum motion. Remarkably, this condition remains unchanged whether or not nondipole effects are included. For later reference, we note that $k_\perp$ determined through Eq.~\eqref{eq:k_full} may vary when the ndSPA or ndSPANE are used. These deviations do not significantly affect the numerical results. The real part of the saddle-point equation leads to
\begin{equation}
t_i = \sqrt{\frac{k_\perp^2+2I_p+p_z^2+\frac{p_z}{c}\left(2\bm{k}_\perp\cdot\bm{A}(t_r)-A^2(t_r) \right)}{\left(1+\frac{p_z}{c}\right)\widetilde{F}^2(t_r)}},
\end{equation}
with the effective field $\widetilde{F}(t_r) = \sqrt{F^2(t_r)-\bm{k}_\perp\cdot\dot{\bm{F}}(t_r)}$. The above expression of $t_i$ explicitly defines an effective Keldysh parameter $\gamma_\mathrm{eff}=\omega t_i$, which should be small in the tunneling regime.

The relation between $\bm{k}_\perp$ and the momentum at the tunnel exit $(\bm{v}_\perp,v_z)$ follows from the Heisenberg's equations of motion for $H_\mathrm{ND}'$ [Eq.~\eqref{eq:H_Total_new}]
\begin{align}
\bm{v}_\perp(t_r, k_\perp, p_z) & = \bm{p}_\perp+\bm{A}(t_r)+\frac{p_z}{c}\bm{A}(t_r) = \bm{k}_\perp+\frac{p_z}{c}\bm{A}(t_r), \\
v_z(t_r, k_\perp, p_z) & = p_z + \frac{1}{c}\left[\bm{p}_\perp\cdot\bm{A}(t_r)+\frac{A^2(t_r)}{2}\right] \nonumber \\
& = p_z + \frac{1}{c}\left[\bm{k}_\perp\cdot\bm{A}(t_r)-\frac{A^2(t_r)}{2}\right].
\end{align}
In addition, the gauge transformation [Eq.~\eqref{eq:GT_all}] implies the frame transformation [Eq.~\eqref{eq:H_Total_new}]
\begin{equation}
\bm{r}_\mathrm{lab} = \bm{r}-\frac{z}{c}\bm{A}(t_r).
\end{equation}
Accordingly, we find for the velocities
\begin{equation}
\bm{v}_\mathrm{lab} = \bm{v}-\frac{v_z}{c}\bm{A}(t_r)+\frac{z}{c}\bm{F}(t_r).
\end{equation}
Using $z\approx0$ at the tunnel exit, we have
\begin{align}
\bm{v}_{\perp,{\rm lab}}(t_r, k_\perp, p_z)&=\bm{v}_\perp-\frac{v_z}{c}\bm{A}(t_r) \approx \bm{k}_\perp, \label{eq:vd_lab} \\
v_{z,{\rm lab}}(t_r, k_\perp, p_z)&=v_z = p_z + \frac{1}{c}\left[\bm{k}_\perp\cdot\bm{A}(t_r)-\frac{A^2(t_r)}{2}\right]. \label{eq:vz_lab}
\end{align}
We note that Eq.~\eqref{eq:vz_lab} enters the expression in Eq.~\eqref{eq:partial} (dropping the subscript ``lab'').

Ionization rate $\tilde{w}_{\rm ndSPA}$ [Eq.~\eqref{eq:WndSPAjac}] is now obtained by evaluating the exponential factor
\begin{align}
2\Im S =& -2I_pt_i - \Re \int_{0}^{t_i} \bigg[ { \bm{p}+\bm{A}(t_r+it) } \nonumber\\
& { +\frac{\hat{\bm{e}}_z}{c}\left(\bm{p}\cdot\bm{A}(t_r+it)+\frac{1}{2}A^2(t_r+it) \right) } \bigg]^2dt \nonumber \\
\approx& -\frac{2}{3\widetilde{F}}\Bigg[ {k_\perp^2+2I_p +\bigg( {p_z} } \nonumber\\
&\left.{ {-\left(\frac{-2\bm{k}_\perp\cdot\bm{A}+A^2 }{2c}+\frac{2I_p+k_\perp^2}{6c}\right) } \bigg)^2} \right]^{\frac{3}{2}},
\end{align}
the preexponential prefactor
\begin{align}
|\ddot{S}|^{-\alpha_Z} \approx& \left|-i\left( 1+\frac{p_z}{c} \right)t_i\widetilde{F}^2 \right|^{-\alpha_Z} \nonumber\\
\approx& \left[\left(k_\perp^2+2I_p \right)\widetilde{F}^2 \right]^{-\frac{\alpha_Z}{2}}\exp\left\{-\frac{\alpha_Z}{2}\frac{p_z}{c} \right\} \nonumber \\
&\times\exp\left\{-\frac{\alpha_Z}{2}\frac{\left(p_z-\frac{-2\bm{k}_\perp\cdot\bm{A}+A^2}{2c} \right)^2}{k_\perp^2+2I_p} \right\},
\end{align}
and the Jacobian
\begin{equation}
\left|\det\frac{\partial(p_x,p_y,p_z)}{\partial(t_r,k_\perp,p_z)}\right| \approx \left|F_d+F(t_r) \right|,
\end{equation}
where $F_d=k_\perp\left[F_x(t_r)F_y'(t_r)-F_x'(t_r)F_y(t_r)\right]/F^2(t_r)$.

Rearranging these expressions, we find
\begin{align}
\tilde{w}_\mathrm{ndSPANE} \approx& |F_d+F(t_r)|\left[\left(k_\perp^2+2I_p\right)\widetilde{F}^2\right]^{-\alpha_Z/2} \nonumber \\
&\times \exp\left\{ {-\frac{2}{3\widetilde{F}}\left[{k_{\perp}^2+2I_p+\left(1+\frac{\alpha_Z\widetilde{F}}{2(k_{\perp}^2+2I_p)^\frac{3}{2}}\right)} \right.}\right.  \nonumber \\
&\times \left.{ \left. {\left(p_z-\langle p_z(t_r,k_\perp)\rangle\right)^2}\right]^\frac{3}{2}} \right \},
\end{align}
where
\begin{align}
\langle p_z(t_r,k_\perp)\rangle= &\frac{2I_p+k_{\perp}^2}{6c}\left[1-\frac{2\alpha_Z\widetilde{F}}{(2I_p+ k_{\perp}^2)^{3/2}}\right] +\frac{-2\bm{k}_\perp \cdot \bm{A} + A^2}{2c} \nonumber \\
\approx &\frac{2I_p+k_{\perp}^2}{6c}\left[1-\frac{2\alpha_ZF}{(2I_p)^{3/2}}\right] +\frac{-2\bm{k}_\perp \cdot \bm{A} + A^2}{2c}.
\label{eq:pz}
\end{align}

Hence, the mean initial linear momentum in the lab frame can be rewritten as
\begin{equation}
\langle v_z(t_r,v_{\perp})\rangle\approx \frac{2I_p+v_{\perp}^2}{6c}\left[1-\frac{2\alpha_ZF(t_r)}{(2I_p)^{3/2} }\right],
\label{eq:vz_Total}
\end{equation}
where we have dropped the subscript ``lab'', and the corresponding asymptotic linear momentum follows as
\begin{align}
\langle p_z(t_r,v_\perp )\rangle &= \langle v_z(t_r,v_\perp)\rangle + \frac{\Delta E}{c}, \nonumber \\
&= \langle v_z(t_r,v_\perp)\rangle + \frac{p^2_\perp-v^2_\perp}{2c}
\label{eq:pz_Total}
\end{align}
with the ponderomotive energy gain $\Delta E$,
\begin{equation}
\frac{\Delta E}{c} = \frac{p_\perp^2-v_\perp^2}{2c}=\frac{-2\bm{v}_\perp \cdot \bm{A} + A^2}{2c}.
\end{equation}

The derivation of analogous expressions for $\langle p_z \rangle$ and $\langle v_z \rangle$ predicted by the various approximate Hamiltonian operators [Eqs.~(\ref{eq:M1}--\ref{eq:F2})] containing selected nondipole terms is given in Appendix~B to E.

The results of the expectation value of the linear momentum in the lab frame for different Hamiltonian are summarized in Table~\ref{tab:vz-pz} in the Appendix, where we present the asymptotic linear momentum transfer $\langle p_z(t_r,v_\perp)\rangle$ [Eq.~\eqref{eq:pz_Total}], which is composed of the initial linear momentum at the tunnel exit $\langle v_z(t_r,v_\perp)\rangle$ [Eq.~\eqref{eq:vz_Total}] and $\Delta E/c$ for the continuum motion after tunneling. From Table~\ref{tab:vz-pz}, it is obvious that
\begin{align}
\langle v_z(t_r,v_\perp)\rangle_{{\rm M}_1} + \langle v_z(t_r,v_\perp)\rangle_{{\rm E}_2} &= \langle v_z(t_r,v_\perp)\rangle, \\
\frac{\Delta E_{{\rm M}_1}}{c} + \frac{\Delta E_{{\rm E}_2}}{c} &= \frac{\Delta E}{c}, \\
\langle v_z(t_r,v_\perp)\rangle_{{\rm F}_1} + \langle v_z(t_r,v_\perp)\rangle_{{\rm F}_2} &= \langle v_z(t_r,v_\perp)\rangle, \\
\frac{\Delta E_{{\rm F}_1}}{c} + \frac{\Delta E_{{\rm F}_2}}{c} &= \frac{\Delta E}{c},
\end{align}
i.e., the sum of the nondipole contributions of the reduced Hamiltonians adds up to that of the full nondipole Hamiltonian within each gauge. This underscores the suitability of the decomposition into reduced Hamiltonians for the observables under consideration.

\section{Subcycle contribution of nondipole terms}

In this section, we quantitatively investigate and illustrate the effects of individual nondipole Hamiltonian terms, most prominently of the magnetic dipole (${\rm M}_1$) and electric quadrupole (${\rm E}_2$) contributions for tunneling ionization for helium. Within SFA helium is effectively represented by a zero-range potential with ionization potential $I_p = 0.903570$. For comparative numerical solutions of the TDSE we use model potential with a Coulomb tail or a short-ranged Yukawa-type potential (see Appendix~A).

For the simulation we use a laser pulse with a vector potential
\begin{equation}
\bm{A}(t) = A_0 \cos^4\left(\frac{\omega t}{2N}\right)
\begin{pmatrix}
\cos(\omega t + \phi_\mathrm{CEP}) \\
\epsilon \sin(\omega t + \phi_\mathrm{CEP})
\end{pmatrix},
\end{equation}
with a number $N$ of cycles, $N = 6$, a carrier-envelope phase $\phi_\mathrm{CEP} = 0$, pulse ellipticity $\epsilon = 0.75$, angular frequency $\omega=0.057$ corresponding to a central wavelength of 800 nm, and the vector potential amplitude $A_0$ corresponding to a laser peak intensity of $5\times10^{14}$ W/cm$^2$. For these laser parameters, the Keldysh parameter is $\gamma\approx0.8$ corresponding to the (moderate) tunneling regime. In all figures that follow, solid lines represent the results calculated using the ndSPA method and the dashed lines stand for those obtained by the ndSPANE method.

\begin{figure}[b]
  \centering \includegraphics[width=\columnwidth]{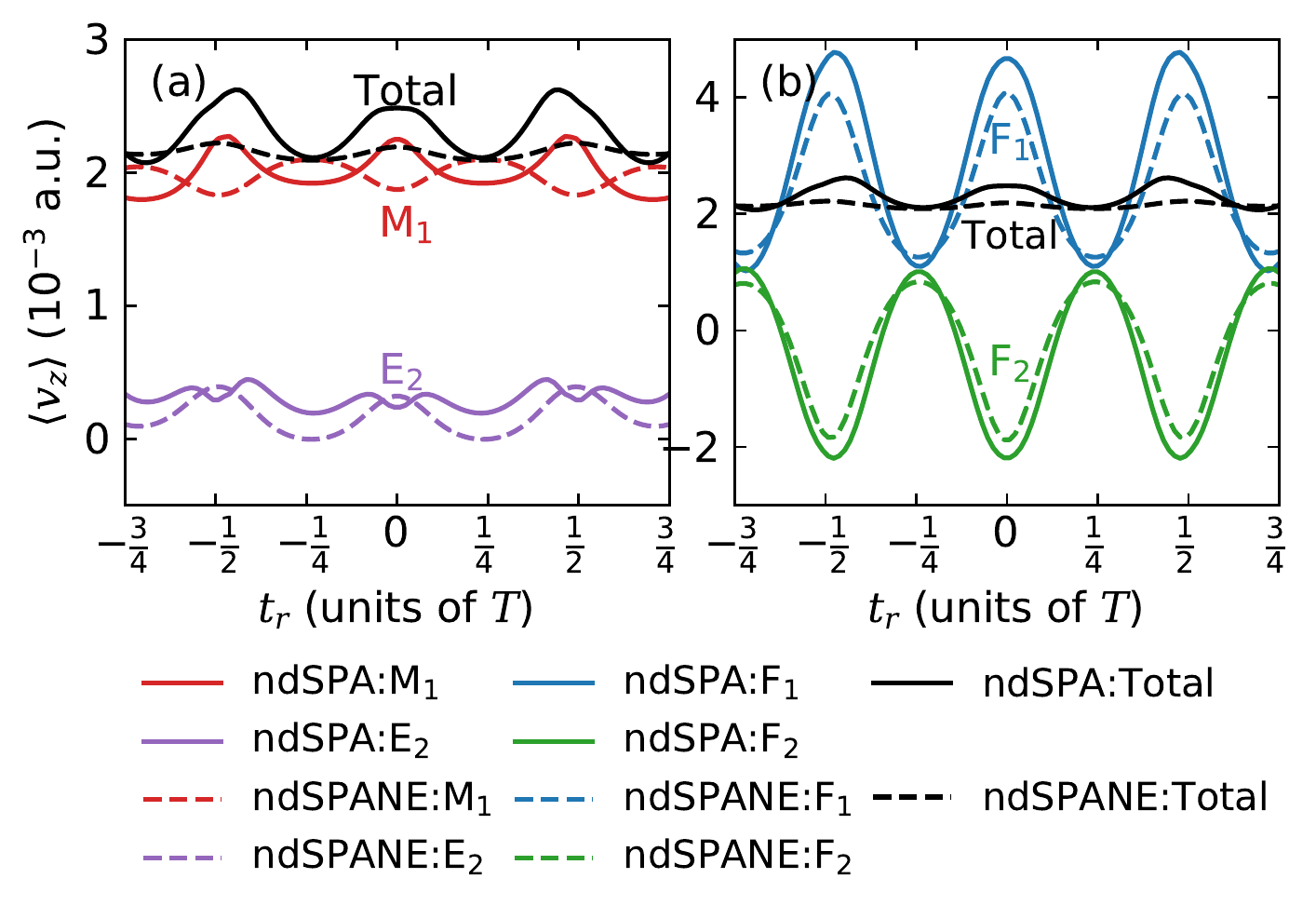}
  \caption{Subcycle time-resolved linear momentum at the tunnel exit $\langle v_z(t_r)\rangle$ based on the Hamiltonian in the multipole gauge [panel (a)] and in the radiation gauge [panel (b)]. Solid curves are calculated using ndSPA and dashed curves are obtained using ndSPANE. The results of individual nondipole Hamiltonian terms are displayed in different colors. Red curve: the M$_1$ term, purple curve: the E$_2$ term, blue curve: the F$_1$ term, green curve: the F$_2$ term, black curve: the total nondipole Hamiltonian.}
  \label{fig:vz}
\end{figure}

Fig.~\ref{fig:vz} shows the subcycle time-resolved linear momentum at the tunnel exit $\langle v_z(t_r)\rangle$ caused by various nondipole Hamiltonian terms. Within the multipole gauge [Fig.~\ref{fig:vz}(a)], we find the remarkable results that the magnetic dipole interaction (M$_1$) strongly dominates the linear momentum transfer during the under-barrier motion while the electric quadrupole interaction (E$_2$) leaves only an almost negligible footprint on the longitudinal momentum at the tunnel exit. For illustrative purposes we also give the corresponding results for the nondipole contributions of the field-linear (F$_1$) and field-quadratic (F$_2$) terms. Here, we find strong oscillations with opposite sign that result in near cancellation of the two contributions. This observation clearly indicates that a decomposition according to the order in coupling with the laser field, unlike the multipole decomposition, does not provide a well-suited starting point for further approximations as both orders are responsible for contributions of comparable magnitude to order $1/c$. This is closely related to the fact that the gauge transformation [Eq.~\eqref{eq:GT_1}] contains field coupling to all orders thereby transforming the Hamiltonian in the radiation gauge [Eq.~\eqref{eq:H_F1F2}] which contains terms of first and second order in the field into the multipole Hamiltonian [Eq.~\eqref{eq:H_E2M1}], which is strictly linear in the field. We also note that the analytic ndSPANE results for all gauges (Fig.~\ref{fig:vz}) slightly differ from the corresponding full ndSPA calculations on the subcycle level due to the neglect of higher-order corrections in the nonadiabatic expansion $\sim t_i^n$ ($n\geqslant3$) [Eq.~\eqref{eq:A}]. Nevertheless, after summation over the corresponding terms within each gauge, the results closely agree with each other.

\begin{figure}[t]
  \centering \includegraphics[width=\columnwidth]{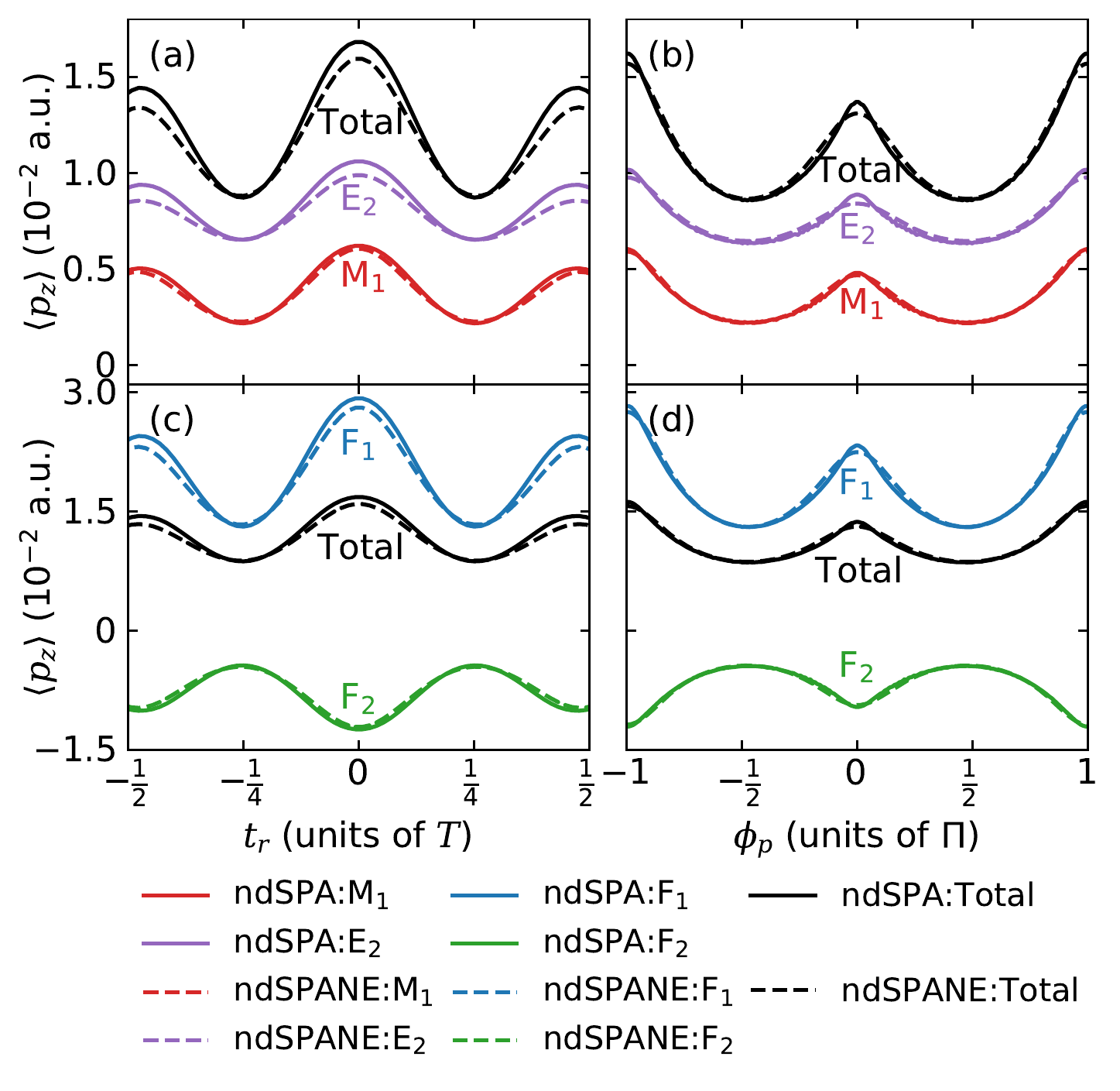}
  \caption{Subcycle asymptotic linear momentum transfer $\langle p_z \rangle$ as a function of the tunneling exit time $t_r$ (left column) and the attoclock offset angle $\phi_p$ (right column) in the multipole gauge (upper row) and in the radiation gauge (lower row). The same legend as in Fig.~\ref{fig:vz} applies.}
  \label{fig:pz}
\end{figure}

Turning now to the asymptotic linear momentum $\langle p_z \rangle$, which is directly experimentally accessible, we display in Fig.~\ref{fig:pz} the dependence on both the release time $t_r$ (left column) and on the attoclock angle $\phi_p=\arctan(p_y/p_x)$ (right column). Most remarkably, while the electric quadrupole term provides only a negligible contribution to the longitudinal momentum transfer during tunneling ionization (Fig.~\ref{fig:vz}), it dominates over the M$_1$ contribution in the post-tunneling continuum momentum transfer [Fig.~\ref{fig:pz}(a)]. Moreover, the asymptotic momentum transfer originates predominantly from the field-driven motion in the continuum. The $2\omega$ subcycle oscillations of the momentum transfer during tunneling mediated by the M$_1$ interaction [Fig.~\ref{fig:vz}(a)] is in phase with the momentum transfer in the continuum, mostly caused by the E$_2$ term [Figs.~\ref{fig:pz}(a,b)]. By contrast, for the decomposition into the field-linear (F$_1$) and field-quadratic (F$_2$) terms of the radiation gauge [Figs.~\ref{fig:pz}(c,d)], the two contribution terms are out of phase by $\pi$. Interestingly, the F$_2$ nondipole term induces a momentum transfer due to the nondipole force $-\left(\bm{A}\cdot\bm{F}\right)/c$ opposed to that of the laser propagation. Time integration of this force leads to a negative momentum transfer of $-A^2(t_r)/2c$. Regardless of partitioning, summation of the respective individual contributions matches the result of the full nondipole Hamiltonian. Here again, the ndSPANE results agree well with full ndSPA calculations. 

\section{Laser parameter dependence of the linear momentum transfer}

In this section, we present results for the laser parameter dependence of the time-integrated linear momentum transfer induced by individual nondipole Hamiltonian terms both at the tunnel exit $\langle v_z\rangle$ and in the asymptotic region $\langle p_z \rangle$ via integration of Eqs.~\eqref{eq:partial} and \eqref{eq:pz_all} over $t_r$ covering the time span of the entire laser pulse.

\begin{figure}[b]
  \centering \includegraphics[width=\columnwidth]{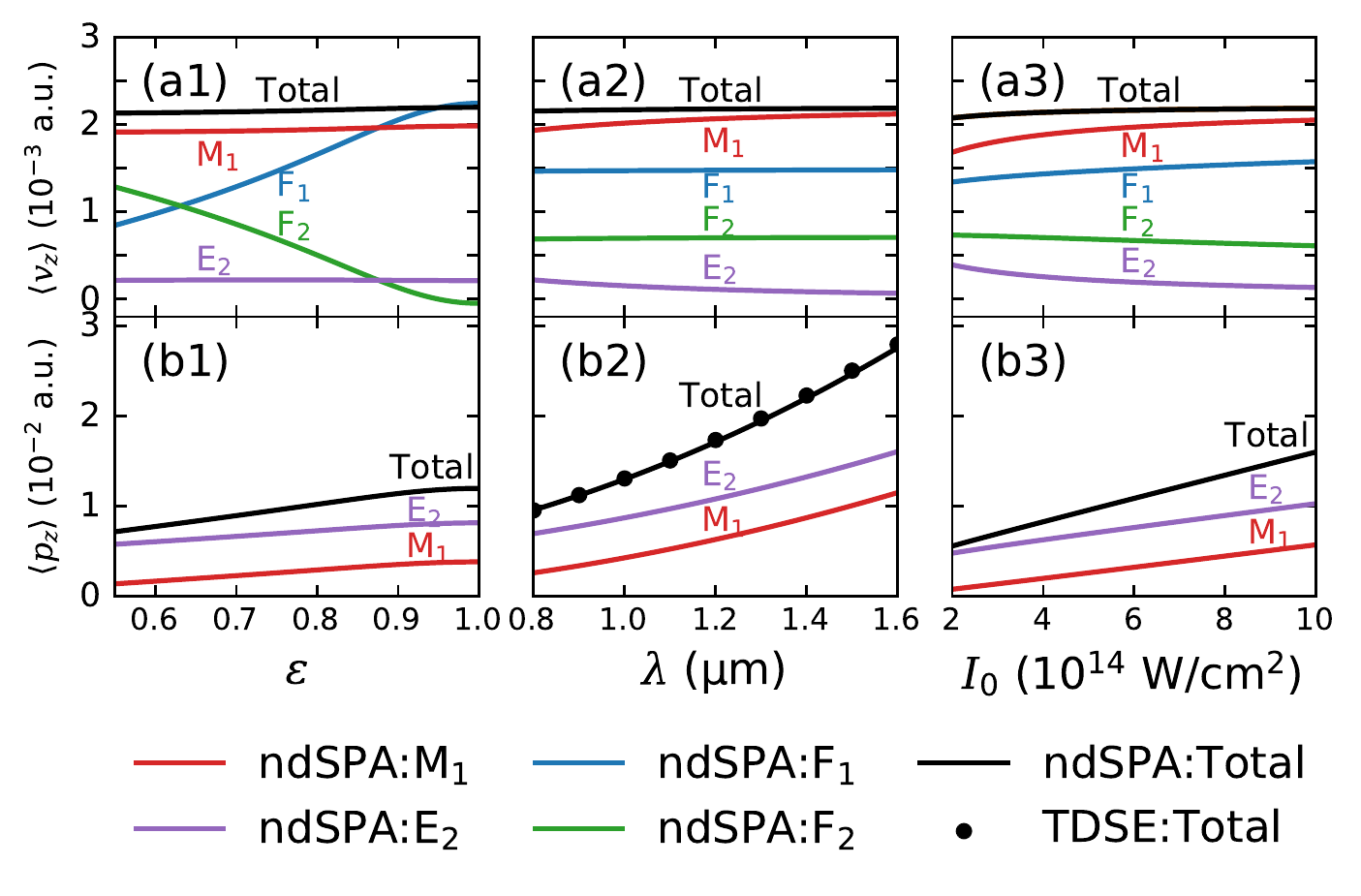}
  \caption{Linear momentum at the tunnel exit $\langle v_z \rangle$ [upper row (a)] and asymptotic linear momentum $\langle p_z \rangle$ [lower row (b)] induced by M$_1$ term (red curves), E$_2$ term (purple curves), F$_1$ term (blue curves), F$_2$ term (green curves), and the full nondipole Hamiltonian (black curves) as a function of laser ellipticity $\epsilon$ (column 1), wavelength $\lambda$ (column 2), and peak intensity $I_0$ (column 3). The black dots in panel (b2) represent the TDSE solution. Laser parameter in column (1): $\lambda=800$ nm, $I_0 = 5\times10^{14}$ W/cm$^2$; column (2): $\epsilon = 0.75$, $I_0 = 5\times10^{14}$ W/cm$^2$; column (3): $\epsilon = 0.75$, $\lambda=800$ nm.}
  \label{fig:Factor}
\end{figure}

Fig.~\ref{fig:Factor}(a) (upper row) shows the dependence of $\langle v_z \rangle$ on the laser ellipticity $\epsilon$ (column 1), wavelength $\lambda$ (column 2), and peak intensity $I_0$ (column 3). Fig.~\ref{fig:Factor}(b) (lower row) shows the corresponding dependences for $\langle p_z \rangle$. In order to demonstrate the level of accuracy reached by the ndSPA for this observable we also present numerical TDSE results for $\langle p_z \rangle$ as a function of $\lambda$ in Fig.~\ref{fig:Factor}(b2). Overall, the under-barrier momentum transfer $\langle v_z \rangle$ is remarkably insensitive to the variation of laser parameters, also the partitioning into M$_1$ and E$_2$ changes only marginally with intensity and wavelength which remains fixed when varying the ellipticity. Throughout, the M$_1$ term strongly dominates over the E$_2$ term. However, when separating the F$_1$ and F$_2$ contributions within the radiation gauge, strong variations as a function of $\epsilon$ can be observed underscoring that this decomposition is not well suited as a starting point for further approximations based on the order of the atom-field interaction.

The asymptotic momentum transfer $\langle p_z \rangle$ (second row) displays a monotonic increase with intensity $I_0$, wavelength $\lambda$, and ellipticity $\epsilon$, in line with the expression
\begin{align}
\langle p_z \rangle &= \langle v_z \rangle + \frac{\langle p_\perp^2\rangle-\langle v_\perp^2\rangle }{2c} = \langle v_z \rangle + \frac{\langle A^2\rangle}{2c} -\frac{\langle\bm{v}_\perp\cdot\bm{A}\rangle}{c} \nonumber \\
&\approx \frac{I_p}{3c} + \frac{\epsilon^2A_0^2}{2c}= \frac{I_p}{3c}+\frac{I_0\lambda^2}{\pi c^4}\frac{\epsilon^2}{1+\epsilon^2},
\end{align}
where we drop the term $-\langle\bm{v}_\perp\cdot\bm{A}\rangle/c$ from the second line. Clearly, the nondipole momentum transfer is proportional to the field intensity $I_0$ and $\lambda^2$. It increases monotonically with ellipticity $\epsilon$. The saturation near $\epsilon\approx1$ is due to the replacement of $\langle A^2\rangle$ by $\epsilon^2A_0^2$ taken at the peak electric field.

For all laser parameters, the E$_2$ contribution dominates over the M$_1$ nondipole contribution to the momentum transfer during the continuum motion and provides, overall, the leading contribution to $\langle p_z \rangle$. The difference between $\langle p_z \rangle$ induced by the E$_2$ term and that by the M$_1$ term remains almost constant with a value of about $2I_p/3c$, originating mainly from their difference during the under-barrier dynamical process (c.f., Table~\ref{tab:vz-pz}).

\section{Conclusions}

In summary, we have systematically investigated the effect of various distinct nondipole terms in the Hamiltonian describing the coupling of the atomic electron to the radiation field. We consider the subcycle time dependence of strong-field tunneling ionization in an attoclock setting. We include terms to leading order in the field retardation $c^{-1}$. In the multipole gauge, the magnetic dipole (M$_1$) interaction and the electric quadrupole (E$_2$) interaction are accounted for. Alternatively, in the radiation gauge, terms linear (F$_1$) and quadratic (F$_2$) in the field strength are considered. The contributions of individual nondipole Hamiltonian terms to the initial linear momentum $\langle v_z \rangle$ at the tunnel exit from the under-barrier motion as well as the asymptotic linear momentum $\langle p_z \rangle$ are studied numerically with the ndSPA method and analytically using the ndSPANE method.

The analysis presented in this work offers novel insights into the nondipole-induced linear momentum on a subcycle time scale. The M$_1$ magnetic dipole effect plays a dominant role in the under-barrier tunneling process while the E$_2$ electric quadrupole effect only leaves a small footprint on the initial linear momentum transfer $\langle v_z \rangle$ at the tunnel exit. This trend is largely independent of the laser parameters. In stark contrast to $\langle v_z \rangle$ at the tunnel exit, the E$_2$ electric quadrupole effect dominates over the M$_1$ magnetic dipole effect in the asymptotic linear momentum $\langle p_z \rangle$. Each of the nondipole terms induce an oscillation with frequency $2\omega$ in the time-resolved electron emission which arises from the coupling of the nondipole and nonadiabatic tunneling effects. With increasing pulse ellipticity, wavelength, and peak intensity, most of individual nondipole contributions increase.

Regardless of the partitioning and gauge, the sum of individual nondipole contributions equals the results of the full nondipole Hamiltonian, implying a clean separation of individual nondipole effects. The identification of the contributions of individual nondipole terms provides novel insights into their importance in different spatiotemporal domains during the strong-field ionization process.

\section*{Acknowledgments}

This work was supported by the National Key R\&D Program of China (Grant No.~2018YFA0306303), the National Natural Science Foundation of China (Grant Nos.~11904103, 92150105, 11834004, 12122404, 11974114), the Austrian Science Fund (Grant Nos.~M2692, W1243), and the Science and Technology Commission of Shanghai Municipality (Grant Nos.~21ZR1420100, 19JC1412200, 19560745900). Numerical computations were in part performed on the ECNU Multifunctional Platform for Innovation (001).

\appendix

\begin{table*}[t]
	\centering
	\renewcommand\arraystretch{1.5}
	\caption{Subcycle time-resolved contributions of individual nondipole Hamiltonian terms to the initial linear momentum $\langle v_z \rangle$ at the tunnel exit and the asymptotic linear momentum $\langle p_z \rangle$. The full nondipole Hamiltonian $H_{\rm ND}$ is reduced to study the influence of the various nondipole contributions, including the magnetic dipole term M$_1$ [Eq.~\eqref{eq:M1}], the electric quadrupole term E$_2$ [Eq.~\eqref{eq:E2}], the field-linear term F$_1$ [Eq.~\eqref{eq:F1}], and the field-quadratic term F$_2$ [Eq.~\eqref{eq:F2}]. The asymptotic linear momentum $\langle p_z \rangle$ relates to the one at the tunnel exit $\langle v_z \rangle$ as $\langle p_z \rangle = \langle v_z \rangle + \Delta E/c$, where $\Delta E/c$ originates from the ponderomotive acceleration in the continuum after tunneling. The initial linear momentum $\langle v_z \rangle$ contains five contributions: the under-barrier motion including the nondipole nonadiabatic coupling, the preexponential prefactor, the reduction factor caused by reducing the full nondipole Hamiltonian, the correction Jacobian of the coordinate transformation, and the correction resulting from the transformation from the respective nondipole frame to the lab frame. The table is summarized from Eqs.~(\ref{eq:vz_Total}--\ref{eq:pz_Total},\ref{eq:vz_M1}--\ref{eq:pz_M1},\ref{eq:vz_E2}--\ref{eq:pz_E2},\ref{eq:vz_F1}--\ref{eq:pz_F1},\ref{eq:vz_F2}--\ref{eq:pz_F2}).}
	\setlength{\tabcolsep}{3.95mm}{
		\centering
		\begin{tabular}{ccccccc}
			\hline \hline
			\multirow{2}{*}{Terms} & \multicolumn{5}{c}{tunneling motion $\langle v_z \rangle$} & \multirow{2}{*}{continuum motion $\frac{\Delta E}{c}$} \\ \cline{2-6}
			~         &under-barrier &exp.~prefactor  &reduction  &Jacobian  &mapping  &  \\ \hline
			Total     &$\frac{2I_p+v_\perp^2}{6c}$  &\multirow{5}{*}{$\times\left[1-\frac{2\alpha_Z F}{\left(2I_p \right)^{3/2}}\right]$}  &$\times1$  &$+0$  &$+0$   &$+\frac{p_\perp^2-v_\perp^2}{2c}$ \\
			~    M$_1$    &$-\frac{2I_p+v_\perp^2}{12c}$  &  &$\times \left(1+\frac{\bm{v}_\perp\cdot\dot{\bm{F}}}{F^2}\right)$  &$-\frac{F}{4c\sqrt{2I_p}}\frac{F}{F_d+F}$  &$+\frac{2I_p+v_\perp^2}{4c}\frac{F^2}{\widetilde{F}^2}$    &$+\frac{p_\perp^2-v_\perp^2}{4c}-\frac{2I_p+v_\perp^2}{4c}\frac{F^2}{\widetilde{F}^2}$  \\
			~    E$_2$    &$\frac{2I_p+v_\perp^2}{4c}$  &  &$\times \left(1+\frac{\bm{v}_\perp\cdot\dot{\bm{F}}}{3F^2}\right)$  &$+\frac{F}{4c\sqrt{2I_p}}\frac{F}{F_d+F}$  &$ - \frac{2I_p+v_\perp^2}{4c}\frac{F^2}{\widetilde{F}^2}$    &$+\frac{p_\perp^2-v_\perp^2}{4c}+\frac{2I_p+v_\perp^2}{4c}\frac{F^2}{\widetilde{F}^2}$  \\
			~    F$_1$    &$\frac{2I_p+v_\perp^2}{6c}$  &  &$\times \left( -\frac{\bm{v}_\perp\cdot \dot{\bm{F}}}{F^2}\right)$  &$-\frac{F}{2c\sqrt{2I_p}}\frac{F}{F_d+F}$  &$-\frac{\bm{v}_\perp\cdot\bm{A}}{c} $    &$-\frac{\bm{v}_\perp\cdot\bm{A}-A^2}{c}$  \\
			~    F$_2$    &$\frac{2I_p+v_\perp^2}{6c}$  &  &$\times \left(1+\frac{\bm{v}_\perp\cdot \dot{\bm{F}}}{F^2}\right)$  &$+\frac{F}{2c\sqrt{2I_p}}\frac{F}{F_d+F}$  &$+\frac{\bm{v}_\perp\cdot\bm{A}}{c} $    &$-\frac{A^2}{2c}$  \\
			\hline \hline
		\end{tabular}}
	\label{tab:vz-pz}
\end{table*}

\section{Influence of the Coulomb potential}

In this paper we focus on the nondipole effects in strong-field ionization as described by variants of the SFA. In order to verify the applicability and approximate validity of the SFA for the observables under consideration, we have also carried out additional full numerical simulations using TDSE. The Coulomb potential of the helium atom is modeled as \nih{\cite{Eckart2018}}
\begin{equation}
V(r) = -\frac{1+e^{-r^2/2}}{\sqrt{r^2+0.14328}}.
\end{equation}
Such model potential of helium captures its essential features such that the asymptotic charge is 1 while the charge seen by the electron at $r=0$ is 2. \nih{The soft-core parameter 0.14328 is applied to avoid the Coulomb singularity that may lead to numerical issues in the practical computations. This numerical model} reproduces the correct ionization potential of helium. We have verfied that this model potential produces very similar numerical results to established model potentials \cite{Tong2005}. In addition, we perform TDSE simulations for a Yukawa-like short-range potential \nih{\cite{Eckart2018}}
\begin{equation}
V(r) = -\frac{1.17822e^{-r/5}+e^{-r^2/2}}{\sqrt{r^2+0.14328}},
\end{equation}
which is expected to be more closely related to the physics incorporated in ndSPA and ndSPANE, which neglect long-range Coulomb effects. From TDSE simulations, we directly obtain the asymptotic linear momentum transfer $\langle p_z \rangle$. To extract the linear momentum at the tunnel exit $\langle v_z \rangle$, on the other hand, we employ, in addition, the backpropagation method \cite{Ni2016,Ni2018a,Ni2018}.

\begin{figure}[t]
	\centering \includegraphics[width=\columnwidth]{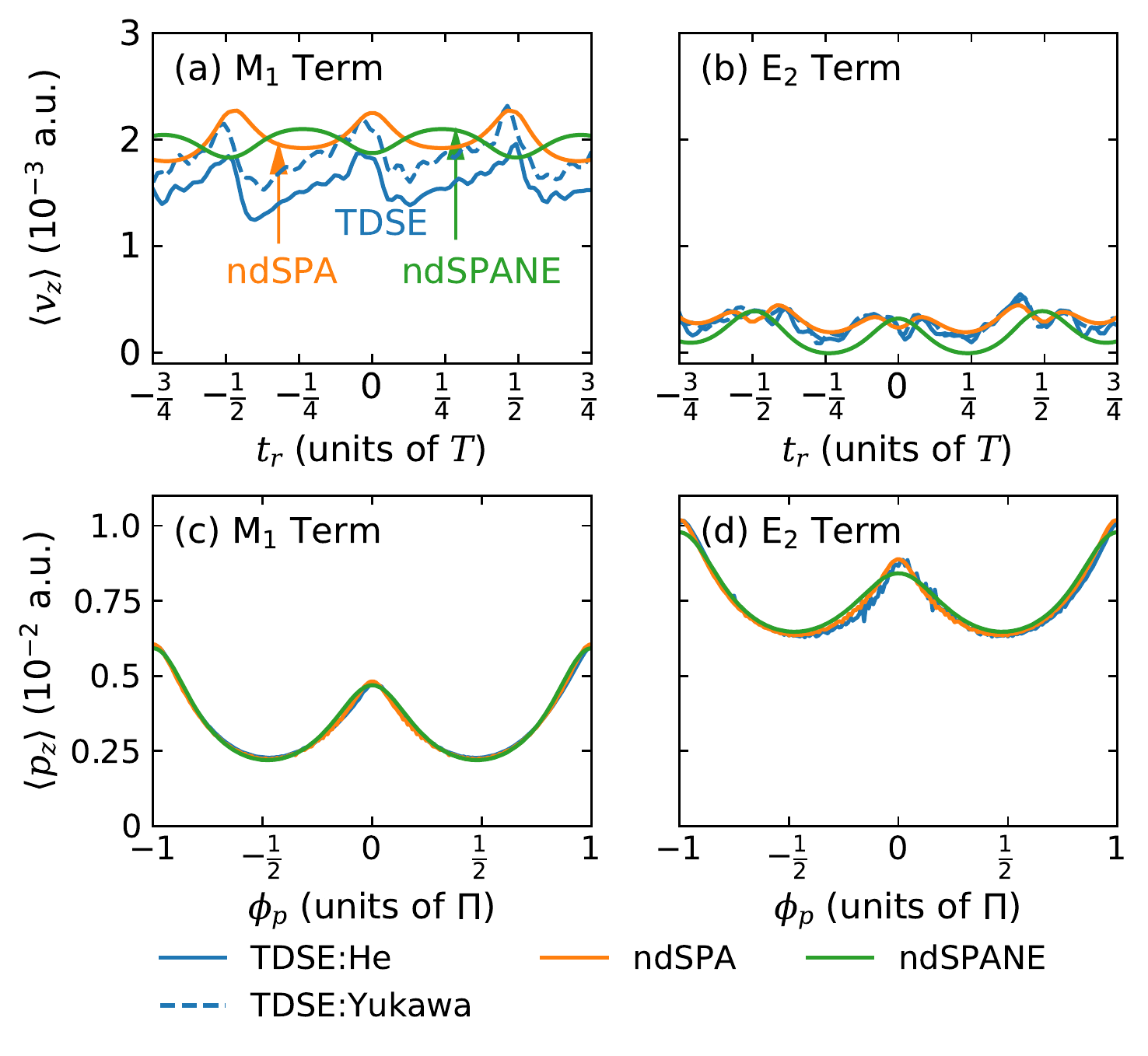}
	\caption{Subcycle time-resolved linear momentum at the tunnel exit $\langle v_z(t_r)\rangle$ (upper row) and asymptotic linear momentum $\langle p_z(\phi_p) \rangle$ (lower row) induced by different nondipole terms calculated by TDSE (blue curves), ndSPA (orange curves), and ndSPANE (green curves).}
	\label{fig:TDSE}
\end{figure}

Shown in Fig.~\ref{fig:TDSE} are the subcycle time-resolved initial linear momentum at the tunnel exit $\langle v_z(t_r)\rangle$ and final longitudinal momentum as a function of the attoclock offset angle $\langle p_z(\phi_p) \rangle$ calculated by TDSE including backpropagation (blue curves), ndSPA (orange curves), and ndSPANE (green curves). In addition, a comparison between the results using the Coulomb potential and the Yukawa potential is also presented. Obviously, the Coulomb potential or, more generally, the atomic potential has only a rather small effect on the nondipole momentum transfer, even when broken down into separate multipole contributions from the M$_1$ and E$_2$ terms. The relatively largest deviation seen is a slight reduction of the momentum transfer at the tunnel exit $\langle v_z \rangle$, which is, however, a minor contribution to total asymptotic momentum transfer $\langle p_z \rangle$.

\section{\texorpdfstring{M$_1$}{M1} term in the multipole gauge}

Similar to the derivation of the nondipole effect of the full Hamiltonian, the contribution of the magnetic dipole term can be studied using the Hamiltonian in the multipole gauge containing only the M$_1$ term
\begin{equation}
	H_{\rm{M_1}} = \frac{1}{2}\bm{p}^2 + V(\bm{r}) + \bm{r}\cdot\bm{F}(t) + \frac{1}{2c}\bm{L}\cdot\bm{B}(t).
	\label{eq:H_M1}
\end{equation}

We carry out two consecutive gauge transformations with $\Lambda_1=-\bm{A}(t)\cdot\bm{r} $ and $\Lambda_2=-\left(z/2c\right)\left[(\bm{p}\cdot\bm{A}(t))+ A^2(t)/2\right]  +\left(p_z/2c\right)\left(\bm{r}\cdot\bm{A}(t)\right)$ to get a new Hamiltonian
\begin{align}
	H'_{\rm{M_1}}=&\frac{1}{2}\left[\bm{p}+\bm{A}(t)+\frac{\hat{\bm{e}}_{z}}{2c}\left(\bm{p}\cdot\bm{A}(t)+ \frac{1}{2}A^2(t) \right)-\frac{p_z}{2c}\bm{A}(t)\right]^2 \nonumber \\
	&+V\left[\bm{r}-\frac{1}{2c}\Big(z\bm{A}(t)-\big(\bm{r}\cdot\bm{A}(t)\big)\hat{\bm{e}}_{z} \Big)\right],
	\label{eq:H_M1_new}
\end{align}
where the corresponding saddle-point equation is written as
\begin{align}
	&\frac{1}{2}\left[\bm{p}+\bm{A}(t_s)+\frac{\hat{\bm{e}}_{z}}{2c}\left(\bm{p}\cdot\bm{A}(t_s)+\frac{1}{2}A^2(t_s)\right)-\frac{p_z}{2c}\bm{A}(t_s)\right]^2 \nonumber \\
	&+I_p=0.
	\label{eq:S_M1}
\end{align}
In the ndSPA method, we define
\begin{equation}
	k_\perp = \left[\bm{p}_\nih{\perp}+\left(1-\frac{p_z}{2c} \right)\Re \bm{A}(t_s) \right]\cdot\frac{\Im A_y(t_s)\hat{\bm{e}}_x - \Im A_x(t_s)\hat{\bm{e}}_y}{\sqrt{\left(\Im A_x(t_s) \right)^2+\left(\Im A_y(t_s) \right)^2}}
\end{equation}
to obtain the linear momentum by evaluating Eq.~\eqref{eq:S_M1}.

In the ndSPANE procedure, we may get
\begin{equation}
	t_i=\sqrt{\frac{k_{\perp}^2+2I_p+p_z^2+\frac{p_z}{c}\left[\bm{k}_\perp\cdot\bm{A}(t_r)-\frac{1}{2}A^2(t_r) \right]     }{\left[1-\frac{p_z}{2c}\left(1+\frac{\bm{k}_\perp\cdot\dot{\bm{F}}(t_r)}{\widetilde{F}^2(t_r)}  \right) \right ] \widetilde{F}^2(t_r) }   },
\end{equation}
where $\bm{k}_\perp = \bm{p}_\perp + (1-p_z/2c)\bm{A}(t_r)$ with $\bm{k}_\perp\cdot\bm{F}(t_r) = 0$. Based on the exponential factor
\begin{align}
	2\Im S=&-2I_pt_i-\Re \int_0^{t_i}  \Bigg [ \bm{p}+\bm{A}(t_r+it)-\frac{p_z}{2c}\bm{A}(t_r+it) \nonumber\\
	&+\frac{\hat{\bm{e}}_{z }}{2c}\left(\bm{p}\cdot\bm{A}(t_r+it)+\frac{1}{2}{A}^2(t_r+it) \right) \Bigg ]^2dt\nonumber \\
	\approx&-\frac{2}{3\widetilde{F}} \Bigg [ k_{\perp}^2 + 2I_p + \Bigg (  p_z    \nonumber \\
	&  -\left(\frac{-\bm{k}_\perp\cdot\bm{A}+\frac{1}{2}A^2 }{2c}-\frac{2I_p+k_{\perp}^2}{12c}\left(1+\frac{\bm{k}_\perp\cdot\dot{\bm{F}}}{\widetilde{F}^2}  \right)\right)  \Bigg)^2 \Bigg ] ^{\frac{3}{2}},
\end{align}
the preexponential prefactor
\begin{align}
	\left|\ddot{S}\right|^{-\alpha_Z}\approx & \left|-i\left[1-\frac{p_z}{2c}\left(1+\frac{\bm{k}_\perp\cdot\dot{\bm{F}}}{\widetilde{F}^2}  \right)  \right]t_i\widetilde{F}^2 \right|^{-\alpha_Z} \nonumber \\
	\approx& \left[\left(k_{\perp}^2+2I_p\right)\widetilde{F}^2\right]^{-\frac{\alpha_Z}{2}} \exp\left\{\frac{\alpha_Z}{2}\frac{p_z}{2c}\left(1+\frac{\bm{k}_\perp\cdot\dot{\bm{F}}}{\widetilde{F}^2}   \right )\right\} \nonumber \\
	&\times\exp\left\{-\frac{\alpha_Z}{2}\frac{\left(p_z-\frac{ -\bm{k}_\perp\cdot\bm{A}+\frac{1}{2}A^2 }{2c}\right)^2}{k_{\perp}^2+2I_p}\right\},
\end{align}
and the Jacobian
\begin{align}
	\left|\mathrm{det}\frac{\partial(p_x,p_y,p_z)}{\partial(t_r,k_\perp,p_z)}\right|& \approx \left|F_d+F(t_r)\right| \left(1-\frac{p_z}{2c}\frac{F(t_r)}{F_d+F(t_r)}\right) \nonumber \\
	&\approx \left|F_d+F(t_r)\right|\exp\left\{-\frac{p_z}{2c}\frac{F(t_r)}{F_d+F(t_r)}  \right \},
\end{align}
the transition rate is written as
\begin{align}
	\tilde{w}_\mathrm{ndSPANE}\approx& \left|F_d+F(t_r)\right|\left[\left(k_{\perp}^2+2I_p\right)\widetilde{F}^2\right]^{-\alpha_Z/2} \nonumber \\
	&\times\exp\left\{ {-\frac{2}{3\widetilde{F}}\left[{k_{\perp}^2+2I_p+\left(1+\frac{\alpha_Z\widetilde{F}}{2(k_{\perp}^2+2I_p)^\frac{3}{2}}\right)} \right.}\right.  \nonumber \\
	&\times\left.{ \left. {\left(p_z-\langle p_z(t_r,k_{\perp})\rangle_{\rm{M_1}}\right)^2}\right]^\frac{3}{2}} \right \},
\end{align}
where
\begin{align}
	\langle p_z(t_r,k_{\perp})\rangle_{\rm{M_1}}=&-\frac{2I_p+k_{\perp}^2}{12c}\left(1+\frac{\bm{k}_\perp\cdot\dot{\bm{F}}(t_r)}{\widetilde{F}^2(t_r)}\right) \left(1-\frac{2\alpha_Z\widetilde{F}(t_r)}{(2I_p+k_\perp^2)^{3/2}}\right) \nonumber \\
	&- \frac{\widetilde{F}(t_r)}{4c\sqrt{2I_p+k_\perp^2}}\frac{F(t_r)}{F_d+F(t_r)}+\frac{-\bm{k}_\perp\cdot\bm{A}+(1/2)A^2}{2c} \nonumber \\
	\approx&-\frac{2I_p+k_{\perp}^2}{12c}\left(1+\frac{\bm{k}_\perp\cdot\dot{\bm{F}}(t_r)}{F^2(t_r)}\right) \left(1-\frac{2\alpha_ZF(t_r)}{(2I_p)^{3/2}}\right) \nonumber \\
	&- \frac{F(t_r)}{4c\sqrt{2I_p}}\frac{F(t_r)}{F_d+F(t_r)}+\frac{-\bm{k}_\perp\cdot\bm{A}+(1/2)A^2}{2c}.
\end{align}

According to the Heisenberg's equations of motion for $H'_{\rm{M_1}}$ [Eq.~\eqref{eq:H_M1_new}],
\begin{align}
\bm{v}_\perp &= \bm{p}_\perp+\bm{A}, \\
v_z &= p_z-\frac{1}{2c}\bm{v}_\perp\cdot\bm{A}+\frac{1}{2c}\left(\bm{p}_\perp\cdot\bm{A}+\frac{1}{2}\bm{A}^2 \right),
\end{align}
and the frame transformation,
\begin{align}
	&\bm{r}_\mathrm{lab} = \bm{r} - \frac{z}{2c}\bm{A} + \frac{\hat{\bm{e}}_z}{2c}\left( \bm{r}\cdot\bm{A} \right),   \\
	&\bm{v}_\mathrm{lab} = \bm{v} - \frac{v_z}{2c}\bm{A}+\frac{z}{2c}\bm{F}+\frac{\hat{\bm{e}}_z}{2c} \left( \bm{v}_\perp\cdot \bm{A} \right)-\frac{\hat{\bm{e}}_z}{2c}\left(\bm{r}\cdot\bm{F} \right),
\end{align}
we can obtain the velocities in the lab frame,
\begin{align}
	\bm{v}_{\perp,{\rm lab}} &= \bm{p}_\perp+\bm{A}-\frac{v_z}{2c}\bm{A}+\frac{z}{2c}\bm{F}, \\
	v_{z,{\rm lab}} &= p_z+\frac{1}{2c}\left(\bm{p}_\perp\cdot\bm{A}+\frac{1}{2}\bm{A}^2 \right)-\frac{1}{2c}\bm{r}\cdot\bm{F}.
\end{align}
At the tunnel exit, using $z\approx 0$ and
\begin{align}
	\bm{r}_0 &= \Re\int_{t_s}^{t_r} \bm{ v_\perp } dt = \Im \int_0^{t_i}\left(1-\frac{p_z}{2c}\right)\bm{A}(t_r+it)dt \nonumber \\
	& =- \frac{1}{2}t_i^2\left(1-\frac{p_z}{2c} \right )\bm{F}(t_r) \nonumber \\
	& \approx -\left(1-\frac{p_z}{2c} \right)\frac{k_\perp^2+2I_p}{2}\frac{\bm{F}}{\widetilde{F}^2},
\end{align}
we have $\bm{v}_{\perp,{\rm lab}} = \bm{k}_\perp$, hence the mean initial linear momentum in the lab frame can be rewritten as
\begin{align}
	\langle v_z(t_r,v_{\perp})\rangle_{\rm{M_1}}\approx&-\frac{2I_p+v_{\perp}^2}{12c}\left(1+\frac{\bm{v}_\perp\cdot\dot{\bm{F}}(t_r)}{F^2(t_r)}\right) \left(1-\frac{2\alpha_ZF(t_r)}{(2I_p)^{3/2}}\right) \nonumber \\
	&- \frac{F(t_r)}{4c\sqrt{2I_p}}\frac{F(t_r)}{F_d+F(t_r)}+ \frac{2I_p+v_\perp^2}{4c}\frac{F^2(t_r)}{\widetilde{F}^2(t_r)},
	\label{eq:vz_M1}
\end{align}
where we have dropped the subscript ``lab'', with the ponderomotive energy gain $\Delta E$,
\begin{equation}
\frac{\Delta E_{\rm M_1}}{c} \approx \frac{p_\perp^2-v_\perp^2}{4c}-\frac{2I_p+v_\perp^2}{4c} \frac{F^2(t_r)}{\widetilde{F}^2(t_r)},
\end{equation}
and the corresponding asymptotic linear momentum
\begin{align}
	\langle p_z(t_r,v_\perp )\rangle_{\rm{M_1}} &= \langle v_z(t_r,v_\perp)\rangle_{\rm{M_1}} +\frac{\Delta E_{\rm{M_1}}}{c}  \nonumber  \\
	&=\langle v_z(t_r,v_\perp)\rangle_{\rm{M_1}} + \frac{ p_\perp^2 - v_\perp^2}{4c} - \frac{2I_p+v_\perp^2}{4c}\frac{ F^2(t_r)}{ \widetilde{F}^2(t_r)}.
	\label{eq:pz_M1}
\end{align}
The results of Eqs.~(\ref{eq:vz_M1}--\ref{eq:pz_M1}) are summarized in Table~\ref{tab:vz-pz}.

\section{\texorpdfstring{E$_2$}{E2} term in the multipole gauge}

The contribution of the electric quadrupole term can be studied using the Hamiltonian in the multipole gauge containing only the E$_2$ term
\begin{equation}
	H_{\rm{E_2}}=\frac{1}{2}{\bm{p}}^2+V(t)+\bm{r}\cdot\bm{F}(t)-\frac{z}{2c}\left[\bm{r}\cdot\dot{\bm{F}}(t)\right]
	\label{eq:H_E2}.
\end{equation}

We carry out two consecutive gauge transformations with gauge phases $\Lambda_1=-\bm{A}(t)\cdot\bm{r}-z\left[\bm{r}\cdot\bm{F}(t)\right]/2c$ and $\Lambda_2=-\left(z/2c\right)\left[(\bm{p}\cdot\bm{A}(t))+ A^2(t)/2\right]  -\left(p_z/2c\right)\left(\bm{r}\cdot\bm{A}(t)\right)$ to get a new Hamiltonian
\begin{align}
	H'_{\rm{E_2}} = &\frac{1}{2}\left[\bm{p}+\bm{A}(t)+\frac{\hat{\bm{e}}_{z}}{2c}\left(\bm{p}\cdot\bm{A}(t)+\frac{1}{2}A^2(t)\right)+\frac{p_z}{2c}\bm{A}(t)\right]^2 \nonumber \\
	&+V\left[\bm{r}-\frac{1}{2c}\Big(z\bm{A}(t)+\big(\bm{r}\cdot\bm{A}(t)\big)\hat{\bm{e}}_{z}\Big)\right],
	\label{eq:H_E2_new}
\end{align}
where the corresponding saddle-point equation is written as
\begin{align}
	&\frac{1}{2}\left[\bm{p}+\bm{A}(t_s)+\frac{\hat{\bm{e}}_{z}}{2c}\left(\bm{p}\cdot\bm{A}(t_s)+\frac{1}{2}A^2(t_s)\right)+\frac{p_z}{2c}\bm{A}(t_s)\right]^2\nonumber \\
	& +I_p=0.
	\label{eq:S_E2}
\end{align}
In the ndSPA method, we define
\begin{equation}
	k_\perp = \left[\bm{p}_\nih{\perp}+\left(1+\frac{p_z}{2c} \right)\Re \bm{A}(t_s) \right]\cdot\frac{\Im A_y(t_s)\hat{\bm{e}}_x - \Im A_x(t_s)\hat{\bm{e}}_y}{\sqrt{\left(\Im A_x(t_s) \right)^2+\left(\Im A_y(t_s) \right)^2}}
\end{equation}
to obtain the linear momentum by evaluating Eq.~\eqref{eq:S_E2}.

In the ndSPANE procedure, we may get
\begin{equation}
	t_i=\sqrt{\frac{k_{\perp}^2+2I_p+p_z^2+\frac{p_z}{c}\left[\bm{k}_\perp\cdot\bm{A}(t_r)-\frac{1}{2}A^2(t_r) \right]}{\left[1+\frac{p_z}{2c}\left(3+\frac{\bm{k}_\perp\cdot\dot{\bm{F}}(t_r)}{\widetilde{F}^2(t_r)}  \right)  \right]\widetilde{F}^2(t_r)  }  },
	\label{eq:ti_E4}
\end{equation}
where $\bm{k}_\perp = \bm{p}_\perp + (1+p_z/2c)\bm{A}(t_r)$, with $\bm{k}_\perp\cdot\bm{F}(t_r) = 0$. Based on the exponential factor
\begin{align}
	2\Im S=&-2I_pt_i-\Re \int_0^{t_i}  \Bigg [ \bm{p}+\bm{A}(t_r+it) +\frac{p_z}{2c}\bm{A}(t_r+it)  \nonumber\\
	&+\frac{\hat{\bm{e}}_{z}}{2c}\left(\bm{p}\cdot\bm{A}(t_r+it)+\frac{1}{2}{A}^2(t_r+it)\right)  \Bigg ]^2dt\nonumber\\
	\approx &-\frac{2}{3\widetilde{F}} \Bigg [ k_{\perp}^2+2I_p + \Bigg ( p_z  \nonumber\\
	&  -\left(\frac{-\bm{k}_\perp\cdot\bm{A}+\frac{1}{2}A^2 }{2c}+\frac{2I_p+k_{\perp}^2}{4c}\left(1+\frac{\bm{k}_\perp\cdot\dot{\bm{F}}}{3\widetilde{F}^2}  \right ) \right )  \Bigg )^2  \Bigg ]^{\frac{3}{2}},
\end{align}
the preexponential prefactor
\begin{align}
	\left |\ddot{S}\right|^{-\alpha_Z} \approx & \left |-i\left[1+\frac{p_z}{2c}\left(3+\frac{\bm{k}_\perp\cdot\dot{\bm{F}}}{\widetilde{F}^2}  \right) \right]t_i\widetilde{F}^2\right |^{-\alpha_Z} \nonumber \\
	\approx&\left[\left(k_{\perp}^2+2I_p\right)\widetilde{F}^2\right]^{-\frac{\alpha_Z}{2}} \exp\left\{-\frac{\alpha_Z}{2}\frac{p_z}{2c}\left(3+\frac{\bm{k}_\perp\cdot\dot{\bm{F}}}{\widetilde{F}^2}    \right)\right\} \nonumber \\
	&\times \exp\left\{-\frac{\alpha_Z}{2}\frac{\left(p_z-\frac{ -\bm{k}_\perp\cdot\bm{A}+\frac{1}{2}A^2 }{2c}\right)^2}{k_{\perp}^2+2I_p}\right\},
\end{align}
and the Jacobian
\begin{align}
	\left|\mathrm{det}\frac{\partial(p_x,p_y,p_z)}{\partial(t_r,k_\perp,p_z)}\right|& \approx \left|F_d+F(t_r)\right|\left(1+\frac{p_z}{2c}\frac{F(t_r)}{F_d+F(t_r)}\right ) \nonumber \\
	&\approx \left|F_d+F(t_r)\right| \exp\left\{ \frac{p_z}{2c}\frac{F(t_r)}{F_d+F(t_r)} \right\},
\end{align}
the transition rate is written as
\begin{align}
	\tilde{w}_\mathrm{ndSPANE}\approx &\left|F_d+F(t_r)\right|\left[\left(k_{\perp}^2+2I_p\right)\widetilde{F}^2\right]^{-\alpha_Z/2} \nonumber \\
	&\times \exp\left\{ {-\frac{2}{3\widetilde{F}}\left[{k_{\perp}^2+2I_p+\left(1+\frac{\alpha_Z\widetilde{F}}{2(k_{\perp}^2+2I_p)^\frac{3}{2}}\right)} \right.}\right.  \nonumber \\
	&\times \left.{ \left. {\left(p_z-\langle p_z(t_r,k_{\perp})\rangle_{\rm{E_2}}\right)^2}\right]^\frac{3}{2}} \right \},
\end{align}
where
\begin{align}
	\langle p_z(t_r,k_{\perp})\rangle_{\rm{E_2}} = &\frac{2I_p+k_{\perp}^2}{4c}\left (1+\frac{\bm{k}_\perp\cdot\dot{\bm{F}}(t_r)}{3\widetilde{F}^2(t_r)}\right)\left (1-\frac{2\alpha_Z\widetilde{F}(t_r)}{(2I_p+k_{\perp}^2)^{3/2}}\right ) \nonumber \\
	&+ \frac{\widetilde{F}(t_r)}{4c\sqrt{2I_p+k_{\perp}^2}}\frac{F(t_r)}{F_d+F(t_r)} + \frac{-\bm{k}_\perp\cdot\bm{A} + (1/2)A^2}{2c} \nonumber \\
	\approx &\frac{2I_p+k_{\perp}^2}{4c}\left (1+\frac{\bm{k}_\perp\cdot\dot{\bm{F}}(t_r)}{3F^2(t_r)}\right)\left (1-\frac{2\alpha_ZF(t_r)}{(2I_p)^{3/2}}\right ) \nonumber \\
	&+ \frac{F(t_r)}{4c\sqrt{2I_p}}\frac{F(t_r)}{F_d+F(t_r)} + \frac{-\bm{k}_\perp\cdot\bm{A} + (1/2)A^2}{2c}.
\end{align}

According to the Heisenberg's equations of motion for $H'_{\rm{E_2}}$ [Eq.~\eqref{eq:H_E2_new}],
\begin{align}
\bm{v}_\perp &= \bm{p}_\perp +\bm{A} +\frac{p_z}{c}\bm{A}, \\
v_z&=p_z+\frac{1}{2c}\bm{v}\cdot\bm{A}+\frac{1}{2c}\left(\bm{p}\cdot\bm{A}+\frac{1}{2}A^2 \right),
\end{align}
and the frame transformation,
\begin{align}
	&\bm{r}_\mathrm{lab} = \bm{r} - \frac{z}{2c}\bm{A} - \frac{\hat{\bm{e}}_z}{2c}\left( \bm{r}\cdot\bm{A} \right),  \\
	&\bm{v}_\mathrm{lab} = \bm{v} - \frac{v_z}{2c}\bm{A}+\frac{z}{2c}\bm{F}-\frac{\hat{\bm{e}}_z}{2c} \left( \bm{v}\cdot \bm{A} \right)+\frac{\hat{\bm{e}}_z}{2c}\left(\bm{r}\cdot\bm{F} \right),
\end{align}
we can obtain the velocities in the lab frame,
\begin{align}
\bm{v}_{\perp,{\rm lab}} &= \bm{p}_\perp +\bm{A}+\frac{v_z}{2c}\bm{A}+\frac{z}{2c}\bm{F}, \\
v_{z,{\rm lab}} &= p_z+\frac{1}{2c}\left(\bm{p}\cdot\bm{A}+\frac{1}{2}A^2 \right) +\frac{1}{2c}\bm{r}\cdot\bm{F}.
\end{align}
At the tunnel exit, using $z\approx 0$ and
\begin{align}
	\bm{r}_0 &= \Re\int_{t_s}^{t_r} \bm{ v_\perp } dt = \Im \int_0^{t_i}\left(1+\frac{p_z}{2c}\right)\bm{A}(t_r+it)dt \nonumber \\
	& =- \frac{1}{2}t_i^2\left(1+\frac{p_z}{2c} \right )\bm{F}(t_r) \nonumber \\
	& \approx -\left(1+\frac{p_z}{2c} \right)\frac{k_\perp^2+2I_p}{2}\frac{\bm{F}}{\widetilde{F}^2},
\end{align}
we have $\bm{v}_{\perp,{\rm lab}}=\bm{k}_\perp$, hence the mean initial linear momentum in the lab frame can be rewritten as
\begin{align}
	\langle v_z(t_r,v_{\perp})\rangle_{\rm{E_2}} \approx &\frac{2I_p+v_{\perp}^2}{4c}\left (1+\frac{\bm{v}_\perp\cdot\dot{\bm{F}}(t_r)}{3F^2(t_r)}\right)\left (1-\frac{2\alpha_ZF(t_r)}{(2I_p)^{3/2}}\right ) \nonumber \\
	&+ \frac{F(t_r)}{4c\sqrt{2I_p}}\frac{F(t_r)}{F_d+F(t_r)} - \frac{2I_p+v_\perp^2}{4c}\frac{F^2(t_r)}{\widetilde{F}^2(t_r)},
	\label{eq:vz_E2}
\end{align}
where we have dropped the subscript ``lab'', with the ponderomotive energy gain $\Delta E$,
\begin{equation}
\frac{\Delta E_{\rm{E_2}}}{c} \approx \frac{p_\perp^2-v_\perp^2}{4c}+\frac{2I_p+v_\perp^2}{4c} \frac{F^2(t_r)} {\widetilde{F}^2(t_r)},
\end{equation}
and the corresponding asymptotic linear momentum
\begin{align}
	\langle p_z(t_r,v_\perp )\rangle_{\rm{E_2}}&= \langle v_z(t_r,v_\perp)\rangle_{\rm{E_2}} + \frac{\Delta E_{\rm{E_2}}}{c}  \nonumber \\
	& = \langle v_z(t_r,v_\perp)\rangle_{\rm{E_2}} + \frac{p_\perp^2-v_\perp^2}{4c} + \frac{2I_p+v_\perp^2}{4c}\frac{F^2(t_r)}{\widetilde{F}^2(t_r)}.
	\label{eq:pz_E2}
\end{align}
These results of Eqs.~(\ref{eq:vz_E2}--\ref{eq:pz_E2}) are summarized in Table~\ref{tab:vz-pz}.

\section{\texorpdfstring{F$_1$}{F1} term in the radiation gauge}

The contribution of the field-linear term can be studied using the Hamiltonian in the radiation gauge containing only the F$_1$ term
\begin{equation}
H_{{\rm F}_1} = \frac{1}{2}\left[\bm{p}+\bm{A}(t)\right]^2 + \frac{z}{c}\bm{F}(t)\cdot\bm{p} + V(r).
\label{eq:H_F1}
\end{equation}

We carry out a gauge transformation with gauge phase $\Lambda = -\left(z/c\right)\left[\bm{p}\cdot\bm{A}(t)\right]$ to get a new Hamiltonian
\begin{equation}
H_{{\rm F}_1}' = \frac{1}{2}\left[\bm{p}+\bm{A}(t)+\frac{\hat{\bm{e}}_z}{c}\bm{p}\cdot\bm{A}(t) \right]^2 + V\left(\bm{r}-\frac{z}{c}\bm{A}(t)\right),
\label{eq:H_F1_new}
\end{equation}
where the corresponding saddle-point equation is written as
\begin{equation}
\frac{1}{2}\left[\bm{p}+\bm{A}(t_s)+\frac{\hat{\bm{e}}_z}{c}\bm{p}\cdot\bm{A}(t_s)\right]^2 + I_p = 0.
\label{eq:S_F1}
\end{equation}
In the ndSPA method, we define
\begin{equation}
k_\perp = \left[\bm{p}_\nih{\perp}+\left(1-\frac{p_z}{c} \right)\Re \bm{A}(t_s)\right]\cdot\frac{\Im A_y(t_s)\hat{\bm{e}}_x - \Im A_x(t_s)\hat{\bm{e}}_y}{\sqrt{\left(\Im A_x(t_s) \right)^2+\left(\Im A_y(t_s) \right)^2}}
\end{equation}
to obtain the linear momentum by evaluating Eq.~\eqref{eq:S_F1}.

In the ndSPANE procedure, we may get
\begin{equation}
t_i = \sqrt{\frac{k_\perp^2+2I_p+p_z^2+\frac{2p_z}{c}\left[2\bm{k}_\perp\cdot\bm{A}(t_r)-A^2(t_r)\right]} {\left[1-\frac{p_z}{c}\frac{\bm{k}_\perp\cdot\dot{\bm{F} }(t_r) }{\widetilde{F}^2(t_r)}\right]\widetilde{F}^2(t_r)}},
\end{equation}
where $\bm{k}_\perp = \bm{p}_\perp + (1-p_z/c)\bm{A}(t_r)$, with $\bm{k}_\perp\cdot\bm{F}(t_r) = 0$. Based on the exponential factor
\begin{align}
2\Im S =& -2I_pt_i - \Re \int_{0}^{t_i}  \bigg[ { \bm{p}+\bm{A}(t_r+it) } \nonumber \\
&  {+\frac{\hat{\bm{e}}_z}{c}\left(\bm{p}\cdot\bm{A}(t_r+it) \right) }  \bigg]^2dt \nonumber \\
\approx& -\frac{2}{3\widetilde{F}}\Bigg [  k_{\perp}^2+2I_p +  \Bigg ( {p_z}   \nonumber\\
&  -\left(\frac{-2\bm{k}_\perp\cdot\bm{A}+A^2 }{c}-\frac{2I_p+k_{\perp}^2}{6c}\frac{\bm{k}_\perp\cdot\dot{\bm{F}}}{\widetilde{F}^2}\right)  \Bigg )^2 \Bigg ]^{\frac{3}{2}},
\end{align}
the preexponential prefactor
\begin{align}
|\ddot{S}|^{-\alpha_Z} \approx& \left|-i\left( 1-\frac{p_z}{c}\frac{\bm{k}_\perp\cdot\dot{\bm{F}}}{\widetilde{F}^2} \right)t_i\widetilde{F}^2 \right| ^{-\alpha_Z}\nonumber\\
\approx& \left[\left(k_\perp^2+2I_p \right)\widetilde{F}^2 \right]^{-\frac{\alpha_Z}{2}}\exp\left\{-\frac{\alpha_Z}{2}\frac{p_z}{c}\frac{\bm{k}_\perp\cdot\dot{\bm{F}}}{\widetilde{F}^2} \right\} \nonumber \\
&\times\exp\left\{-\frac{\alpha_Z}{2}\frac{\left(p_z-\frac{-2\bm{k}_\perp\cdot\bm{A}+A^2}{c} \right)^2}{k_\perp^2+2I_p}  \right\},
\end{align}
and the Jacobian
\begin{align}
\left|\mathrm{det}\frac{\partial(p_x,p_y,p_z)}{\partial(t_r,k_\perp,p_z)}\right|& \approx \left|F_d+F(t_r)\right|\left(1-\frac{p_z}{c}\frac{F(t_r)}{F_d+F(t_r)}\right ) \nonumber \\
&\approx \left|F_d+F(t_r)\right| \exp\left\{ -\frac{p_z}{c}\frac{F(t_r)}{F_d+F(t_r)} \right\}  ,
\end{align}
the transition rate is written as
\begin{align}
\tilde{w}_\mathrm{ndSPANE} \approx& |F_d+F(t_r)|\left[\left(k_{\perp}^2+2I_p\right)\widetilde{F}^2\right]^{-\alpha_Z/2} \nonumber \\
&\times \exp\left\{ {-\frac{2}{3\widetilde{F}}\left[{k_{\perp}^2+2I_p+\left(1+\frac{\alpha_Z\widetilde{F}}{2(k_{\perp}^2+2I_p)^\frac{3}{2}}\right)} \right.}\right.  \nonumber \\
&\times \left.{ \left. {\left(p_z-\langle p_z(t_r,k_{\perp})\rangle_{\rm{F_1}}\right)^2}\right]^\frac{3}{2}} \right \},
\end{align}
where
\begin{align}
\langle p_z(t_r,k_{\perp})\rangle_{\rm{F_1}} = &-\frac{2I_p+k_{\perp}^2}{6c}\frac{\bm{k}_\perp\cdot\dot{\bm{F}}}{\widetilde{F}^2}\left[1-\frac{2\alpha_Z\widetilde{F}}{(2I_p+ k_{\perp}^2)^{3/2}}\right] \nonumber \\
&- \frac{\widetilde{F}}{2c\sqrt{2I_p+k_{\perp}^2}}\frac{F}{F_d+F}  +\frac{-2\bm{k}_\perp\cdot\bm{A}+A^2 }{c} \nonumber \\
\approx &-\frac{2I_p+k_{\perp}^2}{6c}\frac{\bm{k}_\perp\cdot\dot{\bm{F}}}{F^2}\left[1-\frac{2\alpha_ZF}{(2I_p)^{3/2}}\right] \nonumber \\
&- \frac{F}{2c\sqrt{2I_p}}\frac{F}{F_d+F}  +\frac{-2\bm{k}_\perp\cdot\bm{A}+A^2 }{c}.
\end{align}

According to the Heisenberg's equations of motion for $H_{{\rm F}_1}'$ [Eq.~\eqref{eq:H_F1_new}],
\begin{align}
\bm{v}_\perp &= \bm{p}_\perp+\bm{A}+\frac{p_z}{c}\bm{A}, \\
v_z &= p_z+\frac{1}{c}\bm{p}\cdot\bm{A},
\end{align}
and the frame transformation,
\begin{align}
&\bm{r}_\mathrm{lab} = \bm{r}-\frac{z}{c}\bm{A}, \\
&\bm{v}_\mathrm{lab} = \bm{v}-\frac{v_z}{c}\bm{A}+\frac{z}{c}\bm{F},
\end{align}
we can obtain the velocities in the lab frame,
\begin{align}
\bm{v}_{\perp,{\rm lab}} &= \bm{p}_\perp + \bm{A} +\frac{z}{c}\bm{F}, \\
v_{z,{\rm lab}} &= p_z+\frac{1}{c}\bm{p}\cdot\bm{A}.
\end{align}
At the tunnel exit, using $z \approx 0$ we have
\begin{equation}
\bm{v}_{\perp,{\rm lab}}=\bm{p}_\perp + \bm{A} \approx\bm{k}_\perp,
\end{equation}
hence the mean initial linear momentum in the lab frame can be rewritten as
\begin{align}
\langle v_z(t_r,v_{\perp})\rangle_{\rm{F_1}} \approx &-\frac{2I_p+v_{\perp}^2}{6c}\frac{\bm{v}_\perp\cdot\dot{\bm{F}}(t_r)}{F^2(t_r)}\left[1-\frac{2\alpha_ZF(t_r)}{(2I_p)^{3/2}}\right] \nonumber \\
&- \frac{F(t_r)}{2c\sqrt{2I_p}}\frac{F(t_r)}{F_d+F(t_r)}-\frac{\bm{v}_\perp\cdot\bm{A}(t_r)}{c},
\label{eq:vz_F1}
\end{align}
where we have dropped the subscript ``lab'', with the ponderomotive energy gain $\Delta E$,
\begin{equation}
\frac{\Delta E_{\rm F_1}}{c} = \frac{-\bm{v}_\perp \cdot\bm{A}(t_r)+A^2(t_r)}{c},
\end{equation}
and the corresponding asymptotic linear momentum
\begin{align}
\langle p_z(t_r,v_\perp )\rangle_{\rm{F_1}} &= \langle v_z(t_r,v_\perp)\rangle_{\rm{F_1}} + \frac{\Delta E_{\rm{F_1}}}{c} \nonumber \\
&= \langle v_z(t_r,v_\perp)\rangle_{\rm{F_1}}-\frac{\bm{v}_\perp\cdot\bm{A}(t_r)-A^2(t_r)}{c}.
\label{eq:pz_F1}
\end{align}
The results of Eqs.~(\ref{eq:vz_F1}--\ref{eq:pz_F1}) are summarized in Table~\ref{tab:vz-pz}.

\section{\texorpdfstring{F$_2$}{F2} term in the radiation gauge}

The contribution of the field-quadratic term can be studied using the Hamiltonian in the radiation gauge containing only the F$_2$ term
\begin{equation}
H_{{\rm F}_2} = \frac{1}{2}\left[\bm{p}+\bm{A}(t) \right]^2 + \frac{z}{c}\bm{A}(t)\cdot\bm{F}(t) + V(\bm{r}).
\label{eq:H_F2}
\end{equation}

We carry out a gauge transformation with gauge phase $ \Lambda = -\left(z/2c\right)A^2(t)$ to get a new Hamiltonian
\begin{equation}
H_{{\rm F}_2}' = \frac{1}{2}\left[\bm{p}+\bm{A}(t)+\frac{\hat{\bm{e}}_z}{2c}A^2(t) \right]^2 + V(\bm{r}),
\label{eq:H_F2_new}
\end{equation}
where the corresponding saddle-point equation is written as
\begin{equation}
\frac{1}{2}\left[\bm{p}+\bm{A}(t_s)+\frac{\hat{\bm{e}}_z}{2c}A^2(t_s) \right]^2+ I_p = 0.
\label{eq:S_F2}
\end{equation}
In the ndSPA method, we define
\begin{equation}
k_\perp = \left[\bm{p}_\nih{\perp}+\left(1+\frac{p_z}{c} \right)\Re \bm{A}(t_s) \right]\cdot\frac{\Im A_y(t_s)\hat{\bm{e}}_x - \Im A_x(t_s)\hat{\bm{e}}_y}{\sqrt{\left(\Im A_x(t_s) \right)^2+\left(\Im A_y(t_s) \right)^2}}
\end{equation}
to obtain the linear momentum by evaluating Eq.~\eqref{eq:S_F2}.

In the ndSPANE procedure, we may get
\begin{equation}
t_i = \sqrt{\frac{k_\perp^2+2I_p+p_z^2+ \frac{2p_z}{c}\left(-\bm{k}_\perp \cdot\bm{A}(t_r)+\frac{1}{2}A^2(t_r) \right)}{\left[1+\frac{p_z}{c}\left(1+\frac{\bm{k}_\perp\cdot\dot{\bm{F}}(t_r) }{\widetilde{F}^2(t_r)} \right) \right] {\widetilde{F}^2(t_r) } } },
\end{equation}
where $\bm{k}_\perp = \bm{p}_\perp + (1+p_z/c)\bm{A}(t_r)$, with $\bm{k}_\perp\cdot\bm{F}(t_r) = 0$. Based on the exponential factor
\begin{align}
2\Im S = & -2I_pt_i - \Re \int_0^{t_i}\left[\bm{p}+\bm{A}(t_r +it)+\frac{\hat{\bm{e}}_z}{2c}A^2(t_r+it) \right]^2dt \nonumber \\
\approx& -\frac{2}{3\widetilde{F}} \Bigg [ k_{\perp}^2+2I_p + \Bigg ( p_z  \nonumber\\
& -\left(\frac{\bm{k}_\perp\cdot\bm{A}-\frac{1}{2}A^2}{c}+\frac{2I_p+k_{\perp}^2}{6c}\left(1+\frac{\bm{k}_\perp\cdot\dot{\bm{F}}}{\widetilde{F}^2}\right)\right)  \Bigg)^2 \Bigg ]^{\frac{3}{2}},
\end{align}
the preexponential prefactor
\begin{align}
\left |\ddot{S}\right|^{-\alpha_Z} \approx & \left |-i\left[1+\frac{p_z}{c}\left (1+\frac{\bm{k}_\perp\cdot\dot{\bm{F}}}{\widetilde{F}^2}  \right ) \right]t_i\widetilde{F}^2\right |^{-\alpha_Z} \nonumber \\
\approx&\left[\left(k_{\perp}^2+2I_p\right)\widetilde{F}^2\right]^{-\frac{\alpha_Z}{2}} \exp\left\{-\frac{\alpha_Z}{2}\frac{p_z}{c}\left(1+\frac{\bm{k}_\perp\cdot\dot{\bm{F}}}{\widetilde{F}^2}    \right)\right\} \nonumber \\
&\times\exp\left\{-\frac{\alpha_Z}{2}\frac{\left(p_z-\frac{ \bm{k}_\perp\cdot\bm{A}-\frac{1}{2}A^2 }{c}\right)^2}{k_{\perp}^2+2I_p}\right\},
\end{align}
and the Jacobian
\begin{align}
\left|\mathrm{det}\frac{\partial(p_x,p_y,p_z)}{\partial(t_r,k_\perp,p_z)}\right|& \approx \left|F_d+F(t_r)\right|\left(1+\frac{p_z}{c}\frac{F(t_r)}{F_d+F(t_r)}\right ) \nonumber \\
&\approx \left|F_d+F(t_r)\right| \exp\left\{ \frac{p_z}{c}\frac{F(t_r)}{F_d+F(t_r)} \right\}  ,
\end{align}
the transition rate is written as
\begin{align}
\tilde{w}_\mathrm{ndSPANE}\approx &|F_d+F(t_r)|\left[\left(k_{\perp}^2+2I_p\right)\widetilde{F}^2\right]^{-\alpha_Z/2} \nonumber \\
&\times\exp\left\{ {-\frac{2}{3\widetilde{F}}\left[{k_{\perp}^2+2I_p+\left(1+\frac{\alpha_Z\widetilde{F}}{2(k_{\perp}^2+2I_p)^\frac{3}{2}}\right)} \right.}\right.  \nonumber \\
&\times\left.{ \left. {\left(p_z-\langle p_z(t_r,k_{\perp})\rangle_{\rm{F_2}}\right)^2}\right]^\frac{3}{2}} \right \},
\end{align}
where
\begin{align}
\langle p_z(t_r,k_{\perp})\rangle_{\rm{F_2}}=&\frac{2I_p+k_{\perp}^2}{6c}\left(1+\frac{\bm{k}_\perp\cdot\dot{\bm{F}}}{\widetilde{F}^2}\right)\left [1-\frac{2\alpha_Z\widetilde{F}}{(2I_p+ k_{\perp}^2)^{3/2}}\right ] \nonumber \\
&+ \frac{\widetilde{F}}{2c\sqrt{2I_p+k_{\perp}^2}}\frac{F}{F_d+F} +\frac{\bm{k}_\perp\cdot\bm{A}-(1/2)A^2}{c} \nonumber \\
\approx &\frac{2I_p+k_{\perp}^2}{6c}\left(1+\frac{\bm{k}_\perp\cdot\dot{\bm{F}}}{F^2}\right)\left [1-\frac{2\alpha_ZF}{(2I_p)^{3/2}}\right ] \nonumber \\
&+ \frac{F}{2c\sqrt{2I_p}}\frac{F}{F_d+F} +\frac{\bm{k}_\perp\cdot\bm{A}-(1/2)A^2}{c}.
\end{align}

Note that there is no translation to the origin in the Hamiltonian in the present case. According to the Heisenberg's equations of motion for $H_{{\rm F}_2}'$ [Eq.~\eqref{eq:H_F2_new}],
\begin{align}
\bm{v}_{\perp,{\rm lab}} &= \bm{p}_\perp +\bm{A} \approx \bm{k}_\perp, \\
v_{z,{\rm lab}} &= p_z+\frac{1}{2c}A^2,
\end{align}
hence the mean initial linear momentum in the lab frame can be rewritten as
\begin{align}
\langle v_z(t_r,v_{\perp})\rangle_{\rm{F_2}} \approx & \frac{2I_p+v_{\perp}^2}{6c}\left(1+\frac{\bm{v}_\perp\cdot\dot{\bm{F}}(t_r)}{F^2(t_r)}\right)\left [1-\frac{2\alpha_ZF(t_r)}{(2I_p)^{3/2}}\right ] \nonumber \\
&+ \frac{F(t_r)}{2c\sqrt{2I_p}}\frac{F(t_r)}{F_d+F(t_r)}+\frac{\bm{v}_\perp\cdot\bm{A}(t_r)}{c},
\label{eq:vz_F2}
\end{align}
where we have dropped the subscript ``lab'', with the ponderomotive energy gain $\Delta E$,
\begin{equation}
\frac{\Delta E_{\rm F_2}}{c} = -\frac{1}{2c}A^2(t_r),
\end{equation}
and the corresponding asymptotic linear momentum
\begin{align}
\langle p_z(t_r,v_\perp)\rangle_{\rm{F_2}} &=\langle v_z(t_r,v_\perp)\rangle_{\rm{F_2}} + \frac{\Delta E_{\rm{F_2}}}{c} \nonumber \\
& = \langle v_z(t_r,v_\perp)\rangle_{\rm{F_2}} - \frac{1}{2c}A^2(t_r).
\label{eq:pz_F2}
\end{align}
The results of Eqs.~(\ref{eq:vz_F2}--\ref{eq:pz_F2}) are summarized in Table~\ref{tab:vz-pz}.

From Table~\ref{tab:vz-pz}, it is clear that the initial linear momentum $\langle v_z(t_r,v_\perp)\rangle$ can be decomposed into five contributions: the main contribution due to the exponential factor representing the under-barrier motion including the nondipole nonadiabatic coupling, the preexponential prefactor, the reduction factor caused by reducing the full nondipole Hamiltonian, the Jacobian factor introduced by the coordinate transformation $(p_x, p_y, p_z)\rightarrow(t_r, k_\perp, p_z)$, and the frame transformation factor mapping the nondipole frame to the lab frame. $\langle v_z \rangle$ is found in all cases to be modulated by the transverse tunneling momentum $v_\perp^2$, which displays a subcycle variation due to nonadiabatic tunneling effects. Therefore, the interplay between nonadiabatic and nondipole tunneling effects results in a subcycle modulation of the linear momentum transfer at the tunnel exit as well of the asymptotic momentum transfer. The preexponential prefactor, arising partly from the nondipole transition element, is the same for all approximate Hamiltonians. Note that the reduction factor arises from separation of nondipole terms and appears only in the reduced Hamiltonians. Furthermore, the Jacobian and the frame transformations provide additive corrections only for these reduced Hamiltonians.

\begin{table}[tb]
	\centering
	\renewcommand\arraystretch{1.5}
	\caption{Relationship between the initial momentum $\bm{v}_\perp$ and the asymptotic momentum $\bm{p}_\perp$ in the polarization plane under the individual nondipole Hamiltonian terms.}
	\setlength{\tabcolsep}{13.0mm}{
		\begin{center}
		\begin{tabular}{cc}
			\hline
			\hline    Terms      &$\bm{v}_{\perp,\mathrm{lab}}$ \\
			\hline    Total      &$\bm{p}_\perp+\bm{A}(t_r)$ \\
			~   M$_1$  & $\bm{p}_\perp+\left(1-\frac{p_z}{2c}\right)\bm{A}(t_r)$ \\
			~   E$_2$  & $\bm{p}_\perp+\left(1+\frac{p_z}{2c}\right)\bm{A}(t_r)$ \\
			~   F$_1$  & $\bm{p}_\perp+\bm{A}(t_r)$ \\
			~   F$_2$  & $\bm{p}_\perp+\bm{A}(t_r)$ \\
			\hline
			\hline
		\end{tabular}
		\end{center}}
	\label{tab:v}
\end{table}

We note that when we treat different nondipole Hamiltonians, different relationships between the initial and asymptotic transverse momentum arise, which are summarized in Table~\ref{tab:v}.

\nih{\section{Influence of the prefactor}}

\begin{figure}[b]
	\centering \includegraphics[width=\columnwidth]{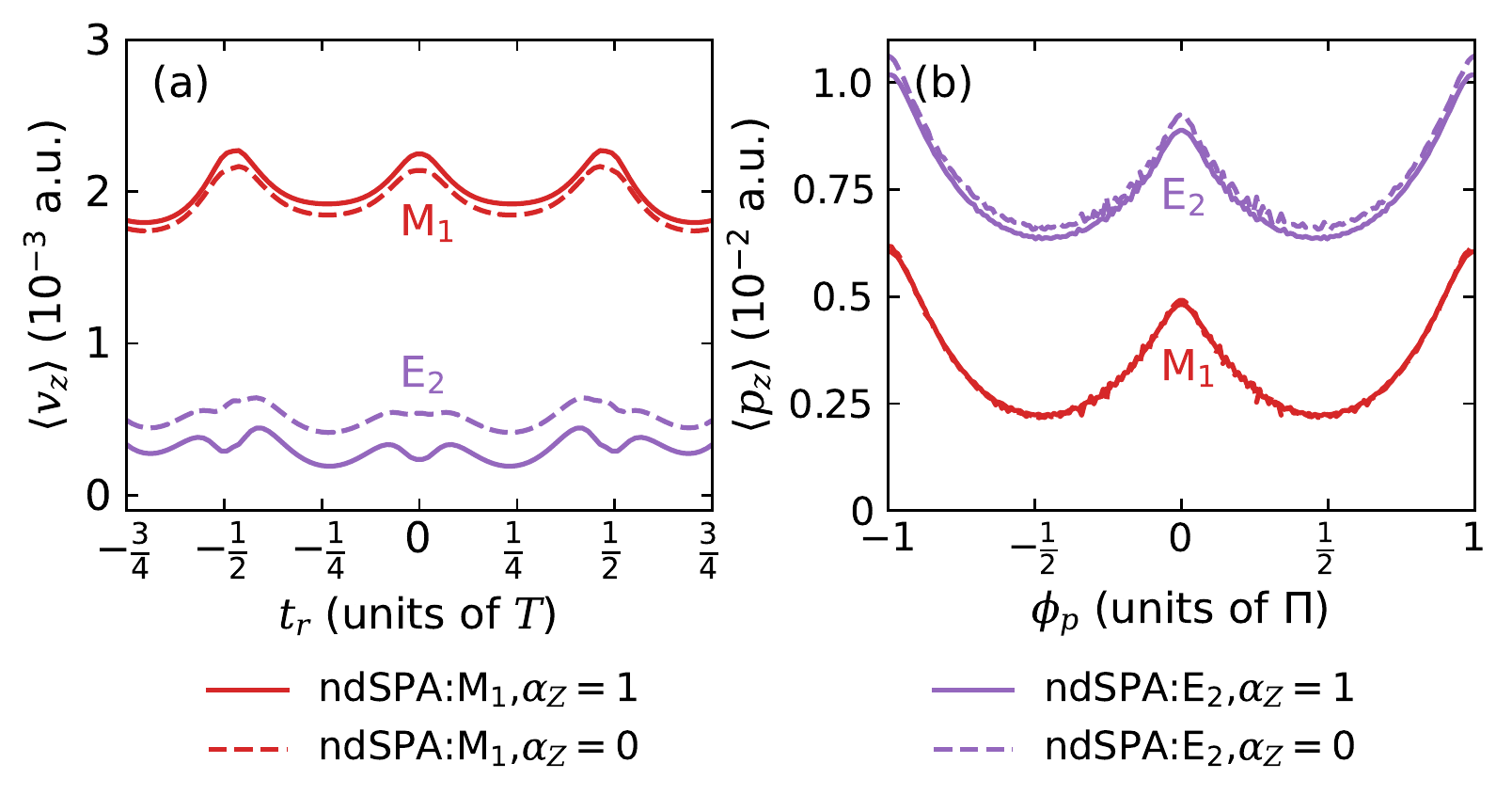}
	\caption{\nih{Subcycle time-resolved linear momentum (a) at the tunnel exit $\langle v_z(t_r)\rangle$ and (b) asymptotic linear momentum $\langle p_z(\phi_p) \rangle$ induced by the M$_1$ term (red curves) and the E$_2$ term (purple curves) calculated by ndSPA with $\alpha_Z=1$ (solid curves) and $\alpha_Z=0$ (dash curves).}}
	\label{fig:alpha}
\end{figure}

\nih{To illustrate the influence of the preexponential prefactor on the linear momentum transfer, we use the ndSPA method to calculate the cases with ($\alpha_Z=1$) and without ($\alpha_Z=0$) the prefactor term in the ionization rate [Eq.~\eqref{eq:rate}], respectively. As shown in Fig.~\ref{fig:alpha}, when the prefactor term is neglected ($\alpha_Z=0$), the linear momentum transfer $\langle v_z \rangle$ and $\langle p_z \rangle$ induced by the M$_1$ term will be slightly reduced, while that of the E$_2$ term will be slightly increased. This can also be concluded from the analytic expressions (see Table~\ref{tab:vz-pz}). For example, the positive influence of the preexponential prefactor in the M$_1$ term results from its multiplication with the negative under-barrier term. Certainly, whether or not the prefactor is considered does not change the conclusions of the present study.}

\end{document}